\documentclass[fleqn,usenatbib]{mnras}

\usepackage[T1]{fontenc}

\DeclareRobustCommand{\VAN}[3]{#2}
\let\VANthebibliography\thebibliography
\def\thebibliography{\DeclareRobustCommand{\VAN}[3]{##3}\VANthebibliography}

\usepackage{graphicx}   
\usepackage{amsmath}    
\usepackage{amssymb}    
\usepackage{subcaption}
\usepackage{pdflscape}

\usepackage{newtxtext,newtxmath}

\newcommand{\snn}{{\sc SuperNNova }}
\newcommand{\ssnn}{{\sc SNN }}
\newcommand{\co}{{\textit{cosmo} }}
\newcommand{\cq}{{\textit{cosmo\_quantile} }}

\title[DES-SN 5YR photometrically identified SNe Ia]{The Dark Energy Survey 5-year photometrically identified Type Ia Supernovae}

\author[Möller et al.]{A. Möller $^{1,2}$\thanks{E-mail: amoller@swin.edu.au} \thanks{Author affiliations are shown in Appendix \ref{appendix:affiliations} }, 
M.~Smith$^{3}$,
M.~Sako$^{4}$,
M.~Sullivan$^{5}$,
M.~Vincenzi$^{6}$,
P.~Wiseman$^{5}$,
P.~Armstrong$^{7}$,
J.~Asorey$^{8}$,
\newauthor
D.~Brout$^{9}$,
D.~Carollo$^{10}$,
T.~M.~Davis$^{11}$,
C.~Frohmaier$^{5,12}$,
L.~Galbany$^{13,14}$,
K.~Glazebrook$^{1}$,
L.~Kelsey$^{5,12}$,
\newauthor
R.~Kessler$^{15,16}$,
G.~F.~Lewis$^{17}$,
C.~Lidman$^{7,18}$,
U.~Malik$^{7}$,
R.C.~Nichol$^{19}$,
D.~Scolnic$^{6}$,
B.~E.~Tucker$^{7}$,
\newauthor
T.~M.~C.~Abbott$^{20}$,
M.~Aguena$^{21}$,
S.~Allam$^{22}$,
J.~Annis$^{22}$,
E.~Bertin$^{23,24}$,
S.~Bocquet$^{25}$,
D.~Brooks$^{26}$,
\newauthor
D.~L.~Burke$^{27,28}$,
A.~Carnero~Rosell$^{21,29,30}$,
M.~Carrasco~Kind$^{31,32}$,
J.~Carretero$^{33}$,
F.~J.~Castander$^{34,35}$,
\newauthor
C.~Conselice$^{36,37}$,
M.~Costanzi$^{38,39,40}$,
M.~Crocce$^{34,35}$,
L.~N.~da Costa$^{21,41}$,
J.~De~Vicente$^{8}$,
S.~Desai$^{42}$,
\newauthor
H.~T.~Diehl$^{22}$,
P.~Doel$^{26}$,
S.~Everett$^{43}$,
I.~Ferrero$^{44}$,
D.~A.~Finley$^{22}$,
B.~Flaugher$^{22}$,
D.~Friedel$^{31}$,
J.~Frieman$^{16,22}$,
\newauthor
J.~Garc\'ia-Bellido$^{45}$,
D.~W.~Gerdes$^{46,47}$,
D.~Gruen$^{25,48}$,
R.~A.~Gruendl$^{31,32}$,
J.~Gschwend$^{21,41}$,
\newauthor
G.~Gutierrez$^{22}$,
K.~Herner$^{22}$,
S.~R.~Hinton$^{11}$,
D.~L.~Hollowood$^{43}$,
K.~Honscheid$^{49,50}$,
D.~J.~James$^{9}$,
K.~Kuehn$^{51,52}$,
\newauthor
N.~Kuropatkin$^{22}$,
O.~Lahav$^{26}$,
M.~March$^{4}$,
J.~L.~Marshall$^{53}$,
F.~Menanteau$^{31,32}$,
R.~Miquel$^{33,54}$,
R.~Morgan$^{55}$,
\newauthor
A.~Palmese$^{56}$,
F.~Paz-Chinch\'{o}n$^{31,57}$,
A.~Pieres$^{21,41}$,
A.~A.~Plazas~Malag\'on$^{58}$,
A.~K.~Romer$^{59}$,
A.~Roodman$^{27,28}$,
\newauthor
E.~Sanchez$^{8}$,
V.~Scarpine$^{22}$,
M.~Schubnell$^{47}$,
S.~Serrano$^{34,35}$,
I.~Sevilla-Noarbe$^{8}$,
E.~Suchyta$^{60}$,
G.~Tarle$^{47}$,
\newauthor
D.~Thomas$^{12}$,
C.~To$^{49}$,
T.~N.~Varga$^{61,62}$
}

\date{Accepted XXX. Received YYY; in original form ZZZ}

\pubyear{2020}

\begin{document}
\label{firstpage}
\pagerange{\pageref{firstpage}--\pageref{lastpage}}
\maketitle

\begin{abstract}
As part of the cosmology analysis using Type Ia Supernovae (SN Ia) in the Dark Energy Survey (DES), we present photometrically identified SN Ia samples using multi-band light-curves and host galaxy redshifts. For this analysis, we use the photometric classification framework \snn trained on realistic DES-like simulations. For reliable classification, we process the DES SN programme (DES-SN) data and introduce improvements to the classifier architecture, obtaining classification accuracies of more than $98$ per cent on simulations. This is the first SN classification to make use of ensemble methods, resulting in more robust samples. Using photometry, host galaxy redshifts, and a classification probability requirement, we identify 1,863 SNe Ia from which we select 1,484 cosmology-grade SNe Ia spanning the redshift range of 0.07 < $z$ < 1.14. We find good agreement between the light-curve properties of the photometrically-selected sample and simulations. Additionally, we create similar SN Ia samples using two types of Bayesian Neural Network classifiers that provide uncertainties on the classification probabilities. We test the feasibility of using these uncertainties as indicators for out-of-distribution candidates and model confidence. Finally, we discuss the implications of photometric samples and classification methods for future surveys such as Vera C. Rubin Observatory Legacy Survey of Space and Time (LSST).
\end{abstract}




\section{Introduction}
To fully exploit the power of current and future time-domain surveys, it is necessary to classify astrophysical objects using only photometry. Surveys such as the Supernova Legacy Survey (SNLS), Sloan Digital Sky Survey (SDSS) SN Survey (SDSS-II), Pan-STARRS (PS1), and the Dark Energy Survey (DES) have discovered thousands of supernovae (SNe) but the majority have not been spectroscopically classified \citep{Astier:2005,Frieman:2008,Sako:2018,Rest:2014,Foley:2018,Bernstein:2011,Smith:2020}. Photometric classification will be particularly crucial for the upcoming Legacy Survey of Space and Time (LSST) at the Vera C. Rubin Observatory, which is expected to discover up to $10^7$ SNe over the next decade \citep{LSST:2009}.

The Dark Energy Survey Supernova programme (DES-SN) obtained photometry of more than $30,000$ candidate SNe over its five years of operation. These include thousands of high-redshift SNe Ia, of which only several hundred have been spectroscopically classified. The first three years of the DES-SN detected and spectroscopically classified 251 SNe Ia \citep{Smith:2020}. Together with low redshift SNe from the Harvard-Smithsonian Center for Astrophysics surveys \citep[CfA3, CfA4;][]{Hicken:2009,Hicken:2012} and the Carnegie Supernova Project \citep[CSP;][]{Contreras:2010,Stritzinger:2011}, these SNe were used to constrain cosmological parameters \citep{Abbott:2019}. The DES-SN candidate sample also contains other types of transients that have been used for astrophysical and cosmological studies: core-collapse SNe \citep{deJaeger:2020}, superluminous SNe \citep[SLSNe;][]{Smith:2018, Angus:2019,Inserra:2021}, rapidly evolving transients \citep{Wiseman:2020a,Pursiainen:2018} and \lq peculiar\rq\ events \citep{Gutierrez:2020,Grayling:2021}. 

To classify SNe without spectroscopy, a number of methods have been developed to classify them using their light-curves, i.e., their observed brightness evolution in different filters. Due to their cosmological use, much work has focused on disentangling SNe Ia from other SN types. The majority have been trained and tested on simulations, with only a handful applied to large SN surveys \citep{Sako:2011,Moller:2016,Muthukrishna:2019,Moller:2020,Villar:2019,Villar:2020}. Several photometric classifiers have been developed and incorporated into the SNIa-cosmology analysis pipeline \textsc{pippin} \citep{Hinton:2020}, including \textsc{snirf} \citep[based on the architecture developed by][]{Dai:2018}, \snn \citep{Moller:2020} and \textsc{scone} \citep{Qu:2021}.

In this work, we use the non-parametric framework \snn \citep[SNN;][]{Moller:2020} to obtain photometrically classified SN Ia samples from DES-SN. \ssnn has several strengths it: (i) requires only photometric information (fluxes and time) for classification, (ii) does not rely on the extraction of features, (iii) can be trained to classify any type of transient event, (iv) can use redshifts to improve accuracy, (v) has been thoroughly tested using simulations, (vi) includes algorithms that assign uncertainties to classification probabilities such as Bayesian Neural Networks (BNNs), and (vii) is already being applied to real survey data, including early light-curve classification in alert streams \citep[{\sc Fink} broker;][]{Moller:2021}.

Photometrically classified SN Ia samples have started to be used in cosmology. First constraints on the cosmic expansion using data from SDSS-II and PS1 have shown the feasibility of using these samples for cosmology and their competitive constraining power on the Dark Energy \citep{Sako:2011,Hlozek:2012,Campbell:2013,Jones:2017,Jones:2018}. Most of these results use the Bayesian Estimation Applied to Multiple Species method \citep[BEAMS;][]{Kunz:2007} and its extension \lq BEAMS with Bias Corrections\rq\ \citep[BBC;][]{Kessler:2017}. These methods incorporate classification probabilities of SNe Ia into the analysis, thus requiring accurate classification probabilities. Recent work estimates the contamination for cosmological constraints in the DES-SN sample using \ssnn at less than $1.4$ per cent \citep{Vincenzi:2021}. Aside from cosmology, photometrically classified samples with \ssnn have also been used to study SN Ia rates \citep{Wiseman:2021}.

This paper is organised as follows: We introduce the DES survey and DES-SN candidate sample in Section~\ref{sec:DES}. In Section~\ref{sec:PC} we present pre-processing needed for accurate classification, {\sc SuperNNova}, realistic simulations, training and classification mechanisms and their metrics. In Section~\ref{sec:PC_wzspe} we select photometrically classified SNe Ia using host galaxy redshift information together with multi-band photometry. We explore the use of BNNs for classification in Section~\ref{section:PC_BNNs}. Finally, in Section~\ref{sec:lsst}, we discuss our results and their implications for future surveys such as LSST.

\section{DES-SN 5-year}
\label{sec:DES}

The Dark Energy Survey (DES) was a 6-year photometric survey that used the Dark Energy Camera \citep[DECam;][]{Flaugher:2015} on the Victor M. Blanco telescope in Chile to survey $5000$\,deg$^2$ of the southern hemisphere. For time-domain science, DES imaged ten $3$-deg$^2$ in the $griz$ filters during the first five years \citep{Abbott:2018}. Eight of these ten fields (X1, X2, E1, E2, C1, C2, S1, and S2) were observed to a single-visit depth of $m\approx 23.5$ mag (\lq shallow fields\rq), and the other two \lq deep fields\rq\ (X3,C3) were observed to a depth of $m\approx24.5$ mag.

\subsection{DES-SN candidate sample} \label{sec:SNcandsample}
Transients were identified using the DES Difference Imaging Pipeline {\sc diffimg} \citep{Kessler:2015} coupled with a machine learning algorithm \citep{Goldstein:2015} to reduce artefacts. A candidate SN is defined from the difference image measurements by requiring at least two detections with a signal-to-noise ratio (SNR) larger than five in any filter. This criteria is designed to remove artefacts and asteroids. 

Each DES-SN candidate was originally associated with a host galaxy using the shallower SVA survey, created from DES Science Verification data. For the DES-SN analysis, we use deep co-adds in \cite{Wiseman:2020}. The major source of host galaxy redshift information was the Australian Dark Energy Survey (OzDES) programme obtaining spectra with the 2dF fibre positioner and AAOmega spectrograph on the 3.9-m Anglo-Australian Telescope \citep{Yuan:2015,Childress:2017,Lidman:2020}. SN hosts in OzDES were observed up to a limiting $r$ magnitude of $\approx 24$. Further details on host galaxy association can be found in \cite{Gupta:2016,Vincenzi:2020}.

For the 31,636 candidates, 29,113 have an identified host and 11,350 have a spectroscopic redshift ($\sim 30$ per cent of the candidates). 

A sub sample of candidates were selected for real-time spectroscopic follow-up observations for classification. For the first 3 years of the survey,the spectroscopically classified sample is presented in \cite{Smith:2020}. In this work, we use for comparison a preliminary spectroscopic sample containing additional classifications from the full 5 years of DES-SN. This sample contains 415 spectroscopically confirmed SNe Ia (including all 251 spectroscopically classified SNe Ia from the DES-SN 3-year analysis), 84 core-collapse SNe, 2 peculiar SNe Ia, 20 SLSNe, 55 AGN, 1 Tidal disruption event (TDE), and 2 M-stars. We highlight that this spectroscopically classified sample is not complete \citep{Kessler:2019} and does not represents the true abundances of different transients in nature.

In this work we use the fluxes and uncertainties obtained from {\sc diffimg} \citep{Kessler:2015} for the DES-SN candidate sample.

\subsection{Filtering multi-season and other transients}\label{sec:multiseason}

The DES-SN 5 year candidate sample contains not only supernovae but also astrophysical events such as fast transients and AGNs. These events, called out-of-distribution (OOD) or anomalies, can be hard to characterise and thus simulate, therefore photometric classifiers are usually not trained to identify them. 

To reject fast, very low SNR transients or transients that have a limited photometric sampling (e.g. transients occurring near the end or beginning of the observing season), we select only transients that have at least 3 nights with a detection that has passed the DES Real/Bogus image classifier \citep{Goldstein:2015}.

To reduce the number of slowly-evolving transients that span several observing seasons or multi-season candidates (e.g. AGNs) and spurious detections we make use of two selection criteria. First, we compute the ratio between number of epochs with detections that pass the Real/Bogus classifier, and the total number of epochs with detections. 
To reject light-curves with long variability periods, we require this ratio to be the same as in the real-time classification pipeline in \cite{Smith:2020}. Second, we remove artefacts and transients that have detections in multiple observing seasons. We note that this cut can remove real supernovae, i.e. multiple SNe very close-by in the same galaxy, and is not 100\% efficient.

With this filtering, the sample is reduced from $31,636$ to $14,070$ candidates. This reduces the number of candidates and their contamination; however, some residual AGN and other types of SNe remain. We find that this sample includes 405 spectroscopically classified SNe Ia (247 of which are in the DES-SN 3yr sample), 83 core-collapse SNe, 2 peculiar SNe Ia, 19 SLSNe, 37 AGN, 1 TDE and 1 M-star.

\subsection{Selection Requirements (cuts)}

We apply a series of selection cuts on both the quality of the light-curves and the quality of the redshifts. A thorough review on these cuts and their impact on systematics can be found in \cite{Vincenzi:2021}.

\subsubsection{Loose selection cuts}\label{sec:selcuts_wz}

We select transients that have redshifts obtained from spectra from either the SN or its host galaxy \citep{Lidman:2020} using quality tags in \cite{Vincenzi:2020}. In this work, we also include lower resolution redshifts from PRIMUS since they are precise enough for photometric classification\footnote{The redshifts from the PRIsm MUlti-object Survey (PRIMUS) were obtained using the Inamori Magellan Areal Camera and Spectrograph camera on the Magellan I Baade 6.5 m telescope \citep{Coil:2011}. They are less accurate and they have a higher rate of catastrophic failure, thus not suitable for cosmological constraints.}. After this selection cut we obtain 6,635 SN candidates.

Furthermore, we restrict these redshifts to be within the range of the SNe Ia expected for DES-SN and thus in our simulations, $z \in [0.05,1.3]$. This cut also removes stars in our catalogues. 

We fit the light-curves using the SALT2 model \citep{Guy:2007}. We require that: (i) at least two filters have at least one observation with SNR larger than 5, (ii) at least one photometric measurement before peak brightness $t_0$, and (iii) at least one photometric point ten days after peak brightness. 

We select a sample of $2,381$ light-curves that satisfy these sampling criteria and have a SALT2 fit that converges and is within SALT2 model boundaries for stretch, $x1 \in [$-$4.9, 4.9]$ and colour $c \in [$-$0.49, 0.49]$. We photometrically classify these candidates in the following. This sample contains a subsample of spectroscopically classified candidates which we will use as a reference: SNe Ia: 366 (DES-SN 3-year: 228), CC 13, SLSN 2, AGN 3. The SALT2 parameters (amplitude, stretch, color) are not used by SNN.

\subsubsection{JLA-like cuts}\label{sec:JLA-like}

We will consider an additional set of cuts after photometric classification based on the criteria in \cite{Vincenzi:2021}. They will only be applied when specified. 

These cuts are designed to select cosmology-grade SNe Ia and are based on those from the Joint Light-curve Analysis: -$3.0<x_1<3.0$, -$0.3<c<0.3$, and $\sigma_{x_1}<1$ and $\sigma_{t0}<2$ \citep{Betoule:2014}. Where $c$, $x_1$, $\sigma_{t0}$, $\sigma_{x_1}$ are estimated using SALT2 and represent colour, stretch and uncertainty on $t_0$ and $x_1$ respectively. These cuts are implemented in SN Ia cosmology analyses to restrict SNIa parameters to the valid model range, and to reject peculiar SNIa. We also use a SALT2 fit probability $>0.001$ selection.

\section{Photometric classification} \label{sec:PC}
We use the photometric classification algorithm \snn ({\sc SNN}) to select SN Ia from the DES-SN 5-year candidate sample that pass loose selection cuts. We introduce pre-processing necessary for accurate photometric classification of our DES-SN 5-year data (Section~\ref{sec:dataprocessing}). We generate realistic simulations of the DES-SN survey to train and test our photometric classification method (Section~\ref{sec:sims}) and the framework \ssnn (Section~\ref{sec:snn}). We evaluate performance and find the best configuration for our framework using small simulations (Section~\ref{sec:snn14X}). We then train optimised models for photometric classification of the DES-SN 5-year sample using larger simulations (Section~\ref{sec:snn26X}). 

\subsection{DES-SN data pre-processing} \label{sec:dataprocessing}
For accurate photometric classification, the simulations used to train the models and the data to be classified should be similar. While light-curve simulations strive to resemble survey data, pre-processing of the survey data is required to assure this.

First, DES-SN data were taken over five consecutive seasons. Each DES season represented about five months of observations per year. SNe last only for months, thus are only detected in a subset of this photometry. In our simulations (see Section~\ref{sec:sims}), supernovae are simulated within a rest-frame time span, e.g. -30 days before to 100 days after peak luminosity. To select an equivalent time window in the DES-SN 5-year data, we first obtain an estimated time of peak brightness ($t_0$) using the SuperNova ANAlysis software \citep[SNANA;][]{Kessler:2009}. This $t_0$ estimate is not obtained using SALT2 \citep{Guy:2007}, but instead based on max flux in region of dense detections to avoid pathological estimates from a single pathological flux in another season. Once the peak has been determined for each light-curve, we select and classify photometric points within an observed time-window around the light-curve peak of $[-30,100]$ days.

Light-curves may contain photometry that has been flagged as flawed. We require that \ssnn discard photometry that is not reliable using the bitmap flag provided by {\sc Source Extractor} \citep{Bertin:1996} and {\sc diffimg} \citep{Kessler:2015}. These photometric outliers are not present in the simulations used to train our photometric classifier. This is in particular important when using normalisation schemes, which will be introduced in Section~\ref{sec:norms}, since they use maximum fluxes to normalise the light-curves. If that maximum flux comes from a bad photometric point, the light-curve will be distorted and therefore classification will not be accurate. This photometry quality criteria reduces the number of photometric measurements by $6\%$ but keeps the number of transients unchanged.

\subsection{Simulations of the DES-SN survey}\label{sec:sims}

\ssnn is used with simulations from the supernova analysis software \citep[snana][]{Kessler:2009} and within the \textsc{pippin} orchestration framework \citep{Hinton:2020}. The simulations incorporate information from DES-SN observations (PSF, sky noise, zero point), with detection efficiencies vs. SNR estimated on fake SNe that were overlaid on images and processed with {\sc diffimg}. Simulations include SNe that have partial light-curves due to season boundaries or observing gaps imitating realistic weather conditions. Detailed information on the inputs necessary to obtain realistic DES-SN simulations can be found in \cite{Kessler:2019}. We also make use of recent updates in the library of simulated host galaxies for DES-SN as introduced by \cite{Vincenzi:2020}. This host galaxy library includes the  dependence of SN rates on galaxy properties such as stellar mass and galaxy star formation rate.    

We simulate a variety of SNe using volumetric rates and input parameters as described in \cite{Vincenzi:2020}. Our simulations are performed over a redshift range $0.05<z<1.3$. These simulations contain normal SNe Ia, peculiar SNe Ia and core-collapse SNe.

Normal SNe Ia are generated using the SALT2 SED model presented in \cite{Guy:2007}, trained for the JLA sample \citep{Betoule:2014} and extended to UV and IR wavelengths \citep{Pierel:2018} to improve the redshift coverage of our simulated SNe. Volumetric rates from \cite{Frohmaier:2019} are used. The intrinsic stretch and colour distributions are taken from \cite{Scolnic:2016} and we use the G10 intrinsic scatter model from \cite{Kessler:2013} based on \cite{Guy:2010}. Peculiar SNe Ia include SN91bg-like \citep{Kessler:2019Plasticc} and SNe Iax \citep{Jha:2017} with models updates in \cite{Vincenzi:2021}. 

We make use of three different core-collapse SN template collections: V19 \citep{Vincenzi:2019}, J17 \citep{Jones:2017} and templates used in the Supernova Photometric Classification Challenge \citep[SPCC;][]{Kessler:2010}. The main differences between these templates include: the number of SNe used to create them, the rates used, and the interpolation methods and wavelength coverage. Detailed information on these templates can be found in \cite{Vincenzi:2019}. 

Our baseline simulations, and used unless specified, are generated using V19 core-collapse SN templates. Relative core-collapse SN rates are given by \cite{Li:2011} updated in \cite{Shivvers:2017} and the total rate is assumed to follow the cosmic star formation history presented in \cite{Madau:2014} normalised by the local SN rate of \cite{Frohmaier:2019}. 

We generate different simulations to train (TRAIN-SIM and a smaller S-TRAIN-SIM for computing efficiency of certain evaluation tasks) and test (TEST-SIM) \ssnn as shown in Table~\ref{tab:sims}. 
For training, after generating the simulation, we randomly trim the simulation to ensure a balanced training sample, with the same number of normal SNe Ia and non-normal Ia (core collapse SNe and peculiar SNe Ia). Volumetric rates guarantee that the mixture of non-Ia SNe is consistent with measured rates. We note that the size of the S-TRAIN-SIM training set is the same as the complete sample used in \cite{Moller:2020}. Having defined our simulated samples we now turn to methods of classifying them.

\begin{table}
\centering
    \caption{Simulations used for training and testing \ssnn. Columns indicate simulation name, approximate number of light-curves generated and number of light-curves when balancing simulations to have the same number of normal Type Ia and other SNe.}
    \label{tab:sims}
\begin{tabular}{lrr}
\hline
              Simulation  &  Number of & Balanced number of \\
               name &  light-curves ($10^6$) & light-curves ($10^6$)\\
\hline
             TRAIN-SIM  & 4.5 & 3.63\\
             S-TRAIN-SIM  & 2.0 & 1.4\\
             TEST-SIM & 0.8 & Not applicable\\
\hline
\end{tabular}

\end{table}

\subsection{\snn (SNN)} \label{sec:snn}
\snn \citep{Moller:2020} is a deep learning framework for light-curve classification. It makes use of fluxes and their measurement uncertainties over time for accurate classification of time-domain candidates. Additional information such as host galaxy redshifts can be included to improve performance.

\begin{figure}
  \centering
\includegraphics[width=\columnwidth]{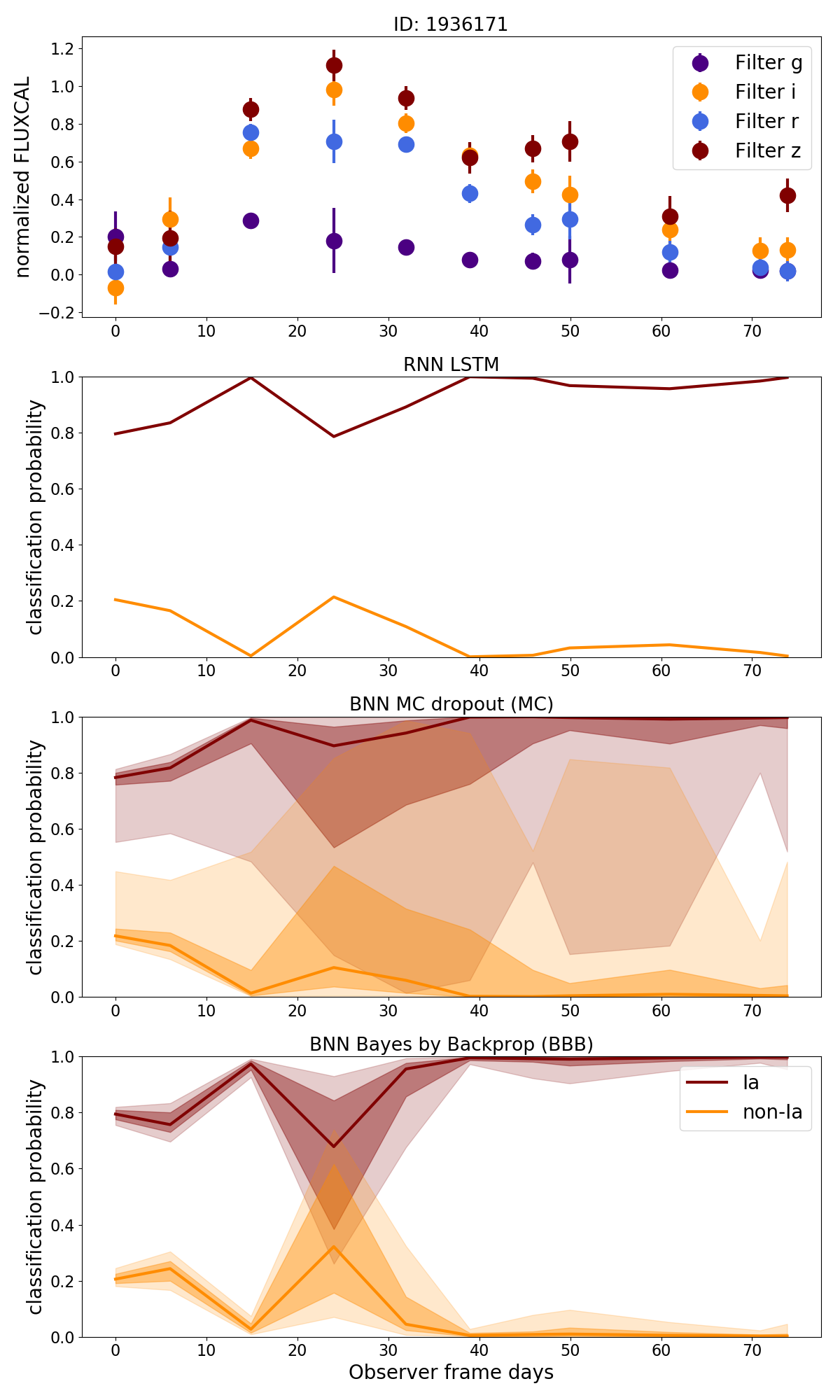}
\caption{\snn ({\sc SNN}) classification for the DES-SN candidate DES17C2hqm at redshift $0.473 \pm 0.001$ using three different neural networks: baseline RNN, BNN MC dropout (MC), and BNN Bayes by Backprop (BBB). All methods were trained with the TRAIN-SIM simulation. Top row shows the SN candidate light-curve from DES (normalised flux with \cq method in band-passes \textit{g,r,i,z}; time in Observer Frame days). Bottom rows shows the classification scores for each method (SN Ia: maroon, non SN Ia: orange). Classification scores use all the data before a given date. The BNN methods provide classification uncertainties (shadowed regions show $68$ per cent and $95$ per cent contours). Each BNN method provides different estimations, this is explored in Section~\ref{sec:PC_BNN_wz_uncertainties}. The large uncertainties in the classification probability represent the lack of confidence in this classification. For this example, uncertainties around days 20--30 are correlated with the lower SNR, while around days 50--60 that correlation is less straight forward to interpret and could be linked to the secondary peak visible in most filters.}
\label{fig:mosaic_lcs_PC}
\end{figure}

\ssnn includes different classification algorithms, such as LSTM\footnote{Long short-term memory \citep[LSTM;][]{Hochreiter:1997})} Recurrent Neural Networks (RNNs) and two approximations for Bayesian Neural Networks (BNNs). We show in Fig.~\ref{fig:mosaic_lcs_PC} the classification probabilities from different methods for a given SN light-curve. These probabilities can be used to select a sample by performing a threshold cut or by weighting the contribution of candidates by their classification score as in the BEAMS and BBC methods \citep{Vincenzi:2021,Kunz:2007,Kessler:2017}.

Light-curve simulations are used to train \ssnn to classify candidates into different classes. For cosmology, it can be trained to accurately classify SNe Ia versus other other kinds of transients. For time-domain astronomy, where brokers are designed to disentangle multiple types of transients, \ssnn can classify subtypes of SNe or transients simultaneously. 

Throughout this work we only perform a binary classification, i.e., a normal SN Ia or a non-Ia SN. Our results are expressed in the form of a prediction of the SN type by using a threshold on the obtained SN Ia probability, $P$, larger than $0.5$.

\subsubsection{SNN normalization schemes: cosmo and cosmo quantile} \label{sec:norms}

Since light-curve fluxes and uncertainties exhibit large variations, \ssnn supports different input data (e.g., fluxes, flux-uncertainties and time steps) and normalisation schemes \citep{Moller:2020}. In previous work, the default was the \textit{global}\footnote{Features, $f$, are log transformed and scaled. The log transform ($f_{\textrm l}$) uses the minimum value of the feature in all band-passes ${\rm min}(f)$ and a constant ($\epsilon$) to centre the distribution at zero as follows: $f_{\textrm l} = \log\left(-{\rm min}(f) + f + \epsilon \right)$. Using the mean and standard deviation of the log transform ($\mu,\sigma$($f_{\textrm l})$), standard scaling is applied: $\hat{f} =  (f_{\textrm l} - \mu(f_{\textrm l})) / \sigma(f_{\textrm l}) $.} normalisation. However, to avoid cosmological bias when using redshifts for classification, it is important to avoid using distance information encoded in the apparent magnitudes.

For classification using redshifts, we introduce two new normalisation schemes in \ssnn that ignore distance information: \co and \textit{cosmo\_quantile}\footnote{Both normalisation schemes are available at: \url{https://github.com/supernnova/SuperNNova}}. In these schemes, for a given light curve, fluxes and their respective uncertainties are normalised by the maximum light-curve flux in any filter (\textit{cosmo}) or the 99th quantile of the flux distribution to avoid normalisation using an outlier (\textit{cosmo\_quantile}). This normalises the fluxes for each light-curve to 1 or near 1, and retains colour and signal-to-noise information for the classification. The normalisation of the time step, given as an input to \ssnn, remains log transformed and displaced to zero as in the \textit{global} normalisation scheme.

To evaluate these new normalisation schemes, we measure the classification accuracy of SN Ia vs non-SN Ia including redshift as an input using simulations from \cite{Moller:2020} since these were the simulations used to benchmark the \ssnn framework. We find that they slightly improve performance with accuracies of $99.33 \pm 0.02$ per cent for both \co and \cq as compared to the $98.43 \pm 0.08$ per cent accuracy of the \textit{global} normalisation scheme using same dataset, redshift information and default settings (seeds and hyper-parameters). In the following analysis, we will use only the \cq norm since it has similar accuracy to \co for the simulations but is more robust against photometry outliers in real data.

\subsection{SNN configuration for performance and robustness}\label{sec:snn14X}
We next study the performance of \ssnn when classifying SNe using photometry and host galaxy redshifts. We also characterise the classification robustness with respect to the training templates, and find the best set of hyper-parameters for our DES-SNIa sample. We use the S-TRAIN-SIM simulations introduced in section~\ref{sec:sims}, for computational efficiency and to compare results with those of \citet{Moller:2020}, to train a classification model. Our simulation was class-balanced (half normal SNe I and half non-Ia SNe) and randomly split in $80$ per cent for training, $10$ per cent for validation and $10$ per cent for metrics evaluation. Uncertainties in the accuracy represent the standard deviation of predictions from five models obtained with different seeds.

Using the default configuration of \ssnn we obtain a classification accuracy of $97.73 \pm 0.04\%$ for the \cq norm. While this accuracy is high, it is $\sim 1\%$ lower than the benchmark in \cite{Moller:2020} for a similar training set size. Since the \ssnn architecture has not been changed, we investigate if this can be attributed to the more complex and realistic DES-SN 5-year simulations in Section~\ref{sec:templates}. We then investigate whether a modified architecture can improve the classification model and thus its accuracy in Section~\ref{sec:hp}. We highlight that \ssnn does not reach its peak performance when trained using the smaller S-TRAIN-SIMS. Thus, larger simulations are needed to improve the model performance.

\subsubsection{Templates impact on performance} \label{sec:templates}
Here, we study how the set of templates used to generate the training simulation impacts the metrics of our classification algorithm. We train different models using simulations that are similar in size (equivalent to S-TRAIN-SIM) but are generated by replacing a subset of templates from the original configuration. Obtained accuracies are shown in Table~\ref{tab:snn_templates}.


\begin{table}
    \caption{Classification accuracies for models trained by replacing a subset of templates from the original configuration in Section~\ref{sec:sims}.}
    \label{tab:snn_templates}
\begin{tabular}{ll}
\hline
Changed template &         accuracy  \\
\hline
JLA instead of extended SNIa model & $97.96\pm 0.05 $  \\
without peculiar SNe Ia & $98.21 \pm 0.01 $  \\
J17 instead of V19 core-collapse model & $98.06\pm 0.07 $  \\
SPCC templates instead of V19 core-collapse model & $98.59\pm0.02$ \\
\hline
\end{tabular}
\end{table}

Models trained with SPCC and J17 templates obtain higher accuracies than those trained with V19 templates. This is consistent with the accuracy decrease of our present model when compared to that of \cite{Moller:2020}. This is evidence of the more complex classification task with the updated simulations. We highlight that V19 uses a large variety of core-collapse templates with greater diversity than previous core-collapse models, J17 and SPCC. From these, SPCC has the fewest number of non-Ia templates and thus less diversity. SPCC templates were used in \cite{Moller:2020} simulations. The impact of changes like using the JLA SALT2 model is less. This shows that the complexity of the classification task increases largely with the updated and more diverse core-collapse SN population in the V19 templates and the inclusion of peculiar SNe Ia.

We thus attribute the decrease on accuracy to the more complex task of disentangling SNe Ia from core-collapse and peculiar SNe Ia generated with updated templates.

\subsubsection{Hyper-parameters}\label{sec:hp}
We investigate whether network hyper-parameters could be modified to improve performance \citep[for a list of available hyper-parameters, see][]{Moller:2020}. We train our models using using $20$ per cent of the S-TRAIN-SIM simulations ($280,033$ light-curves). We modify: batch size ($128, 512$), dropout ($0.05, 0.1, 0.2$), bidirectional (True, False), hidden dimensions ($ 32,64,128$), number of layers ($2,3,4$), two learning policies (cyclic and non-cyclic) and different cyclic phases when using cyclic ($[5, 10, 15], [20,40,60]$). We find that the accuracy in different configurations varies up to $\approx 2 \%$. We find that deeper (3 or 4 layers) and wider networks (up to 64 hidden dimensions) result in the biggest changes to the accuracy. This reflects the increasing complexity of the classification task with updated SN templates. Our chosen configuration for S-TRAIN-SIM is: batch size 512, dropout 0.05, bidirectional network, 64 hidden dimensions, 4 layers, and non-cyclic learning policy. Using the whole S-TRAIN-SIM dataset with this new configuration, the classification accuracy rises to $98.10 \pm 0.06$ per cent. 

\subsection{SNN trained models for DES-SN 5-year analysis}\label{sec:snn26X}
In the following we use \ssnn models trained with a larger dataset to improve classification accuracy, TRAIN-SIM, and the best configuration of \ssnn found in the previous section. We increase the batch size to 1024 for efficient resource allocation. The larger simulation and optimised hyper-parameters provide a better classification accuracy with accuracies above $98\%$ as shown in Table~\ref{tab:snnsims_wz}. Accuracies are computed with a balanced test set, where half of the candidates are SNe Ia and half are non-Ia SNe. 

To evaluate the accuracy, efficiency and purity of our photometric samples, we estimate the performance of our models in the independent TEST-SIM. This simulation is not balanced and thus reflect the relative rate between SN types. We present performance metrics for different levels of selection cuts in Table~\ref{table:metrics_wcuts}. We highlight that we provide the balanced accuracy which shows that after the JLA-like cuts, the remaining non-Ia SNe are harder to disentangle. A thorough analysis on systematics linked to this classification method can be found in \cite{Vincenzi:2021}.

In this work, the traditional classification method is named "single model". This method represents classifications done using probabilities obtained from one SNN trained model with a single seed. In the following, we provide a mean value and uncertainty on the metric or classified sample of the "single model" method by taking the probabilities obtained with 5 models trained with different seeds. These probabilities are then used to compute the mean and standard deviation of the metrics listed in Table~\ref{tab:snnsims_wz}.

\subsubsection{Ensemble methods}\label{sec:ensemble}

For cosmology, we aim to have a classification method that is not highly sensitive to statistical fluctuations in the model and training dataset. In ML, ensemble methods have been shown obtain more robust predictions \citep{Dietterich:2020,Lakshminarayanan:2016} and have been introduced for regression in astronomy \citep{Kim:2015, CarrascoKind:2014}. To produce ensemble classifications, predictions from multiple models are combined. This can be viewed as a mechanism of Bayesian marginalisation \citep{Wilson:2020,Izmailov:2021} and an alternative to Bayesian Neural Networks using Variational Inference explored in Section~\ref{section:PC_BNNs}.

We explore two possible ensemble methods: "probability averaging" and "target averaging".  Probability averaging uses the probability scores and averages them to select light-curves that are above the 0.5 probability threshold of being SN Ia. The "target average" method averages the predictions and selects the most common one. Uncertainties are computed using the standard deviation of the metric for three different sets of five models with different seeds.

We find that ensemble methods increase the accuracy and purity $\approx 0.1\%$ from just using one model prediction, or "single model", as can be seen in Table~\ref{tab:snnsims_wz}. We find a $99.4\%$ overlap between photometrically selected Type Ia SNe using both the ensemble and single model methods. In the following, we will use the "probability average" from different models as our ensemble method.

Each ensemble in this work is obtained using the predictions of 5 models trained with different seeds, also called an "ensemble set". To study the performance of ensemble methods, we compute metrics using the output of 3 ensemble sets, quoting their mean and standard deviation.

\begin{table}
    \caption{SNN Baseline Performance vs. method on TRAIN-SIM without cuts. The chosen method in this work is the Ensemble (probability average) and is highlighted in bold.}
    \label{tab:snnsims_wz}
\begin{tabular}{llll}
\hline
              method &         balanced accuracy &       efficiency &           purity \\
\hline
\multicolumn {4} {c}{ \textit{cosmo} }\\
\hline

                single model &  $98.33\pm 0.01$ &  $98.65\pm 0.05$ &  $98.03\pm 0.06$ \\
                ensemble (target av.) &  $98.43\pm 0.02$ &   $98.81\pm 0.02$ &  $98.08\pm 0.02$ \\
ensemble (prob. av.)  &  $98.45\pm 0.01$ &  $98.80\pm 0.02$ &  $98.11\pm 0.02$ \\
\hline
\multicolumn {4} {c}{ \textit{cosmo\_quantile }}\\
\hline
               single model &  $98.35\pm 0.01$ &  $98.68\pm 0.07$ &  $98.03\pm 0.05$ \\
ensemble (target av.) &  $98.45\pm 0.00$\footnotemark &  $98.84\pm 0.02$ &  $98.09\pm 0.01$ \\
 \textbf{Ensemble (prob. av.)} &  $\mathbf{98.46\pm 0.01}$ &  $\mathbf{98.83\pm 0.03}$ &   $\mathbf{98.10\pm 0.03}$ \\
\hline
\end{tabular}
\end{table}
\footnotetext{We provide only two-significant figures. The uncertainties are negligible and less than 0.005.\label{footnot:significant}}

\begin{table}
\caption{SNN Baseline Performance vs. method on TEST-SIM with loose selection and JLA-like cuts.}\label{table:metrics_wcuts}
\begin{tabular}{llll}
\hline               method &        balanced accuracy &      efficiency &            purity \\

\hline
\multicolumn {4} {c}{ {with loose selection cuts}}\\
\hline
single model & $98.61 \pm 0.03$ & $99.61 \pm 0.02$ &   $99.43 \pm 0.02$ \\
ensemble (prob. av.) & $98.69 \pm 0.01$ & $99.68 \pm 0.01$ & $99.45 \pm 0.00$\textsuperscript{\ref{footnot:significant}} \\
\hline
\multicolumn {4} {c}{ {+ JLA-like cuts}}\\
\hline
single model & $98.26 \pm 0.06$ &   $99.81 \pm 0.01$ &    $99.7 \pm 0.01$ \\
\textbf{Ensemble (prob. av.)} & $\mathbf{98.36 \pm 0.01}$ & $\mathbf{99.86 \pm 0.01}$ & $\mathbf{99.71 \pm 0.00}$\textsuperscript{\ref{footnot:significant}} \\

\hline
\end{tabular}
\end{table}

\subsubsection{Generalisation} \label{sec:generalization}
In this Section we verify the ability of our trained models to classify data generated using different simulation templates. This is called generalisation and showcases the adaptation of our SNN models to new unseen data.

We evaluate the accuracy of our models when trained with simulations generated using SNe Ia, peculiar SNe and the V19 core-collapse templates but applied to simulations generated using other core-collapse templates such as J17 or SPCC. We observe a decrease of $<0.5 \%$ in accuracy, which shows that our V19 trained models generalise well to other templates of core-collapse SNe.

We find that ensemble methods such as \textit{probability average} reduces the loss in accuracy due to changes in the data by $0.2\%$ relative to the single model. This is expected as ensemble methods are usually more robust and thus generalise better than single models.

\subsection{Bayesian Neural Networks (BNNs)}
In scientific analyses using machine-learning outputs, it is important to evaluate the reliability of a model’s predictions, expressed through uncertainties. Uncertainties can be divided into: \textit{Aleatoric}, usually linked to measurement uncertainties (e.g. noise or other effects of data acquisition); \textit{Epistemic} or model uncertainty, which encompasses uncertainties in the training set and NN architecture.

In this section we introduce Bayesian Neural Networks (BNNs) which are a promising method to provide uncertainties reflecting the model's confidence on the prediction.

To compute uncertainties, we obtain different classification probabilities for a given input and evaluate their variance. In NNs this is equivalent to finding a posterior distribution of weights. Typically, this posterior distribution is intractable for deep neural networks, thus different methods have been developed to approximate it. A review on BNNs, approximation methods and their use in astronomy can be found in \cite{Charnock:2020}. 

In this Section we use two BNN implementations approximating the posterior distribution of weights: MC dropout \citep{Gal:2015} and Bayes by Backprop \citep{Fortunato:2017}. MC dropout (MC in the following) provides a Bayesian interpretation by using the same dropout mask at the different NN layers including the recurrent ones (each time step). Bayes by Backprop (BBB in the following) learns a posterior distribution of weights which can then be sampled. Both methods have been previously implemented and tested on simulations in \ssnn \citep{Moller:2020}.

\subsubsection{BNN classification probabilities and uncertainties}

For both methods, to obtain the classification probability distribution, we sample the predictions from our BNN 50 times. This sampling number is also known as as the number of \textit{inference samples}, $n_s$. In the following we compute the classification probability, $P_i$ for a given light-curve, $\mathbf{x}_i$ as the mean of sampled probabilities:
\begin{equation}\label{eq:BNN_probability}
P_i = \frac{1}{n_s} \left( \sum_{j=1}^{n_s} p_{j}({\mathbf{x}_i})\right)
\end{equation}
where $j \in [1,n_s]$ is the index of inference samples, $p_{j}({\mathbf{x}_i})$ is the $j^{th}$ sample of the classification probability distribution for the light-curve $\mathbf{x}_i$.

We compute the classification probability uncertainty for a given light-curve $\mathbf{x}_i$ as the standard deviation of sampled probabilities:
\begin{equation}\label{eq:uncertainty}
\widehat \sigma_i = \frac{1}{n_s}\sqrt{ \sum_{j=1}^{n_s} \left(p_{j}({\mathbf{x}_i}) - P_i \right)^2}
\end{equation}
where $j \in [1,n_s]$ is the index of inference samples, $p_{j}({\mathbf{x}_i})$ is a classification probability for the given light-curve $\mathbf{x}_i$ for each inference sample $j$, and $P_i$ is given by Equation~\ref{eq:BNN_probability}. 

\subsubsection{BNN trained models} \label{sec:BNN_probs}
Using the TRAIN-SIM simulations we train the two Bayesian models, MC and BBB, for light-curve classification with host galaxy redshifts. Both methods obtain high classification accuracies for the ensemble probability average method, $98.33\pm 0.01$ and $98.11\pm 0.01$ for MC dropout and BBB respectively. Balanced accuracies are slightly lower than the ensemble method in Table~\ref{tab:snnsims_wz}. These may be improved by adjusting of the hyper-parameters. We choose to keep the current configuration and focus on the behaviour of the classification uncertainties. 

Traditionally, BNNs are not used in ensembles, combining predictions by different models. To do so, ideally, the probability distributions for each model in the ensemble set should be concatenated into a "joint probability distribution". Then, the ensemble classification probability would be computed using Equation~\ref{eq:BNN_probability} sampling $n_s$ times the "joint probability distribution". However, this can be computationally expensive. Using TEST-SIM simulations, we find that averaging the mean probability obtained for each model in the ensemble set is a close approximation of the one obtained using "joint probability distribution". We find that the differences between probabilities using the approximation and the "joint distribution" are centred at $0.00 \pm 0.01$ and accuracies change by less than $0.1\%$. We use this approximation in the following for computational efficiency. 

\begin{table}
\caption{Performance metrics of BNNs evaluated using TEST-SIM simulations with JLA-like cuts. These simulations are indicative of the expected purity and efficiency of our photometrically classified samples.}\label{table:snnsims_bnn_wz}
\begin{tabular}{llll}
               method &        accuracy &      efficiency &            purity \\
\hline
\multicolumn {4} {c}{ {MC with JLA-like cuts}}\\
\hline
single model & $98.01 \pm 0.03$ & $98.41 \pm 0.03$ & $97.63 \pm 0.07$ \\
ensemble (prob. av.) &$98.11 \pm 0.01$ & $98.51 \pm 0.06$ & $97.73 \pm 0.05$ \\
\hline
\multicolumn {4} {c}{ {BBB with JLA-like cuts}}\\
\hline
single model &$98.01 \pm 0.03$ & $98.41 \pm 0.03$ & $97.63 \pm 0.07$\\
{ensemble (prob. av.)} & $98.11 \pm 0.01$ & $98.51 \pm 0.06$ & $97.73 \pm 0.05$ \\
\end{tabular}
\end{table}

We also test approximating ensemble uncertainties as the sum of uncertainties from each model in the ensemble set assuming the covariance between models is zero. We find on average that the uncertainties obtained with this approximation and from the "joint probability distribution" are similar. However, we note that the approximation for the BBB method has a larger dispersion than the one for the MC method. We will evaluate the potential use of BNN classification uncertainties in Section~\ref{sec:vro_bnn}. 

We use TEST-SIM to evaluate the expected metrics for our photometrically classified samples with JLA-like cuts in Table~\ref{table:snnsims_bnn_wz}. The samples obtained with BNNs have less than 3\% contamination but that is higher than our Baseline DES-SNIa samples with JLA-like cuts. BNN performance could be eventually be improved with a different network configuration and initialisation. However, for comparison we keep this architecture for the analysis in Section~\ref{section:PC_BNNs}.

\section{DES-SN 5-year photometrically classified SNe Ia} \label{sec:PC_wzspe}

In this Section, we photometrically classify DES-SN 5-year candidates with host spectroscopic redshifts using our baseline RNN trained in Section~\ref{sec:snn26X}. 

First, we classify candidates that pass loose cuts using \ssnn trained with host galaxy redshifts in Section~\ref{sec:PC_norm}. We further constrain the sample using JLA-like cuts and visual inspection in Section~\ref{sec:towards_cosmo}. We discuss possible contamination of this sample in Section~\ref{sec:contamination} and its classification efficiency in Section~\ref{sec:PCwz_classeff}. We summarise the properties of the baseline photometrically classified SN Ia sample with JLA-like cuts in Section~\ref{sec:PCwz_properties}.

\subsection{Photometric classification} \label{sec:PC_norm}

We use our baseline RNN model to select photometrically classified SNe Ia. We show the number of selected light-curves in Table~\ref{tab:PC_wz} and their overlap with spectroscopic SN samples defined in Section~\ref{sec:DES}. 

As shown in Sections~\ref{sec:ensemble} and \ref{sec:generalization}, ensemble methods provide more robust predictions than single model methods. We select our Baseline SNe Ia sample using the "probability average" method and the \cq norm. This normalisation is more robust towards photometry outliers present in our analysis. We note that the overlap between \co and \cq probability average sample is larger than $98\%$ and between \cq probability average and single model samples is larger than $99\%$.

Our Baseline DES-SNIa sample contains 1,863 photometrically identified SNe Ia passing loose selection cuts. In this sample, twelve spectroscopically classified SNe Ia are not selected, representing less than $1\%$ of the photometric sample. We do not find a particular redshift or SALT2 parameter preference for these lost SNe Ia. Visual inspection reveals some light-curves have variable quality photometry which could contribute to the mis-classification.

The baseline sample with loose selection cuts can be used to study astrophysical properties of SNe Ia like correlations with their host galaxies, diversity and rates. In the following, we further constrain this sample with cosmology-grade cuts as in \cite{Vincenzi:2021}.

\begin{table}
    \centering
    \caption{DES-SNIa photometric samples with different selection cuts. In the last row, we define the \textbf{Baseline DES-SNIa sample} using a single ensemble set probability threshold. Columns indicate: the number of photometrically selected SNe Ia and the number of spectroscopically classified SNe Ia contained in that sample.}
    \label{tab:PC_wz}
\begin{tabular}{lllll}

               &     \multicolumn{2}{c}{loose selection cuts} &  \multicolumn{2}{c}{{+JLA-like cuts}} \\
              method &     photo Ia &   spec Ia & {photo Ia} & {spec Ia} \\
\hline
        single model & $ 1861^{+17}_{-13} $ & $ 353^{+3}_{-3} $ & $     1478^{+10}_{-11} $ & $       320^{+3}_{-2} $\\
ensemble (prob. av.) & $   1867^{+3}_{-4} $ & $ 354^{+1}_{-0} $ & $       1482^{+2}_{-2} $ & $       321^{+1}_{-0} $\\
\\
\hline
\textbf{Baseline DES-SNIa sample} & \textbf{1863} & \textbf{354} & \textbf{1484} & \textbf{321} \\
 \hline

\end{tabular}

\end{table}

\subsection{Cuts towards a cosmology sample (JLA-like)} \label{sec:towards_cosmo}

We further constrain our sample by applying selection cuts based on SALT2 light-curve fits and redshift quality.

First, we implement additional requirements on the fitted SALT2 parameters of the photometrically selected SNe Ia. As in \cite{Vincenzi:2021}, we implement the JLA-like SALT2 cuts from the Joint Light-curve Analysis \citep{Betoule:2014} introduced in Section~\ref{sec:JLA-like}. Second, we select only candidates which have a high-precision spectroscopic redshift. We eliminate those candidates that have redshifts provided by PRIMUS since the spectra are of lower-resolution, more prone to catastrophic failures and not high-quality enough for cosmology analysis. 

The results of these cuts in the photometrically selected samples are shown in Table~\ref{tab:PC_wz}. We highlight that the JLA-like cuts reduce the scatter in the number of SNe, as can be seen by the reduced standard deviation in the Table when compared to the sample without JLA-like cuts. We obtain a Baseline DES-SNIa sample with JLA-like cuts of $1,484$ photometrically classified SNe Ia. The missing spectroscopic SNe Ia are found to be redder in average and at all redshifts with a median around 0.5. 

A summary of the selection criteria used to obtain this sample can be found in Table~\ref{tab:summary_selcuts_PC}. General properties of these samples are further studied in Section~\ref{sec:PCwz_properties}. 


\begin{table*}
    \centering
    \caption{Effect of the selection cuts on the candidate sample. We show results for the shallow and deep fields, as well as the total number. Note that some events belong to both shallow and deep fields due to field overlap. Columns show the cut, the number of selected candidates, the number of spectroscopic SN Ia in the sample and the Section where the sample is described.}
    \label{tab:summary_selcuts_PC}
\begin{tabular}{llllllll}            
cut & \multicolumn{2}{c}{shallow} & \multicolumn{2}{c}{deep} &  \multicolumn{2}{c}{total} &  \\
 & selected & spec Ia & selected & spec Ia & selected & spec Ia & section\\       
\hline 
DES-SN 5-year candidate sample &  29203 &  415 &7500 & 93 &31636 & 415 & \ref{sec:SNcandsample}\\              
Multi-season & 13868 &  405 &4428 & 88 &14070 & 405 & \ref{sec:multiseason}\\ 
Redshifts in 0.05<z<1.3 &    6556 &  401 &1812 & 85 & 6590 & 401 & \ref{sec:selcuts_wz}\\
SALT2 loose selection &  2380 &  366 & 698 & 77 & 2381 & 366 & \ref{sec:selcuts_wz}\\
RNN>0.5 (\textbf{Baseline DES-SNIa}) &  1863 &  354 & 502 & 76 & 1863 & 354 & \ref{sec:PC_norm}\\
JLA-like (\textbf{Baseline DES-SNIa JLA}) & 1484 &  321 & 408 & 73 & 1484 & 321 & \ref{sec:towards_cosmo} 
\end{tabular}
\end{table*}

\subsection{Contamination}\label{sec:contamination}

As shown in \cite{Vincenzi:2021} and in Table~\ref{table:metrics_wcuts} contamination from core-collapse and peculiar SNe in a SNN classified sample with quality cuts is expected to be less than $1\%$. This estimate was obtained using SN simulations containing various types of core-collapse and peculiar SNe. We inspect the Baseline DES-SNIa sample with JLA-like cuts obtained in the previous section and do not find any spectroscopically identified core-collapse or peculiar SNe. We note that spectroscopic samples are not complete and DES-SN follow-up preferentially targeted suspected Type Ia SNe.

In this section, we explore a different type of potential contaminant, "out-of-distribution" candidates such as AGNs and other unknown transients. These candidates can be erroneously classified since they are not present in the simulated training sample and thus we do not know how \ssnn classifies them.

We find no spectroscopically identified AGN, SLSNe or other SN spectral types in our Baseline DES-SNIa sample but 5 candidates with host spectra showing AGN features. We find that DES16E2nb, DES16X1ext, DES13X3dbe are displaced by more than $1''$ from the centre of the galaxy (additionally DES16E2nb is a spectroscopic Type Ia SN) and the other two candidates are displaced between $0.5''$ and $1''$. At these separations, the light-curves from these candidates are not dominated by the AGN which we confirm by inspection of the light-curves. Therefore we keep these photometrically selected SNe Ia in our Baseline DES-SNIa sample. 

We also perform visual inspection of the light-curves in the Baseline DES sample. We find 3 candidates that can be visually tagged as multi-season visually: DES16E2nb a spectroscopic SN Ia with close by AGN, DES16C3nd two SN Ia in a galaxy \citep{Scolnic:2020}, DES14E2rpm a spectroscopic SN Ia with a fake SN inserted at the same coordinates \citep[fakes were inserted to evaluate the detection efficiency in DES-SN images, see][]{Brout:2019}.We keep all these candidates since they are real supernovae with fake or other SN light-curves that do not overlap.

Photometrically classified Type Ia SNe samples are expected to have some level of contamination from core-collapse and peculiar SNe and possibly by other transients. For the Baseline DES-SNIa sample in this work we find no clear evidence of contamination from core-collapse and peculiar SNe or long-term variables such as AGNs.

\begin{figure*}
    \centering
    \includegraphics[width=\textwidth]{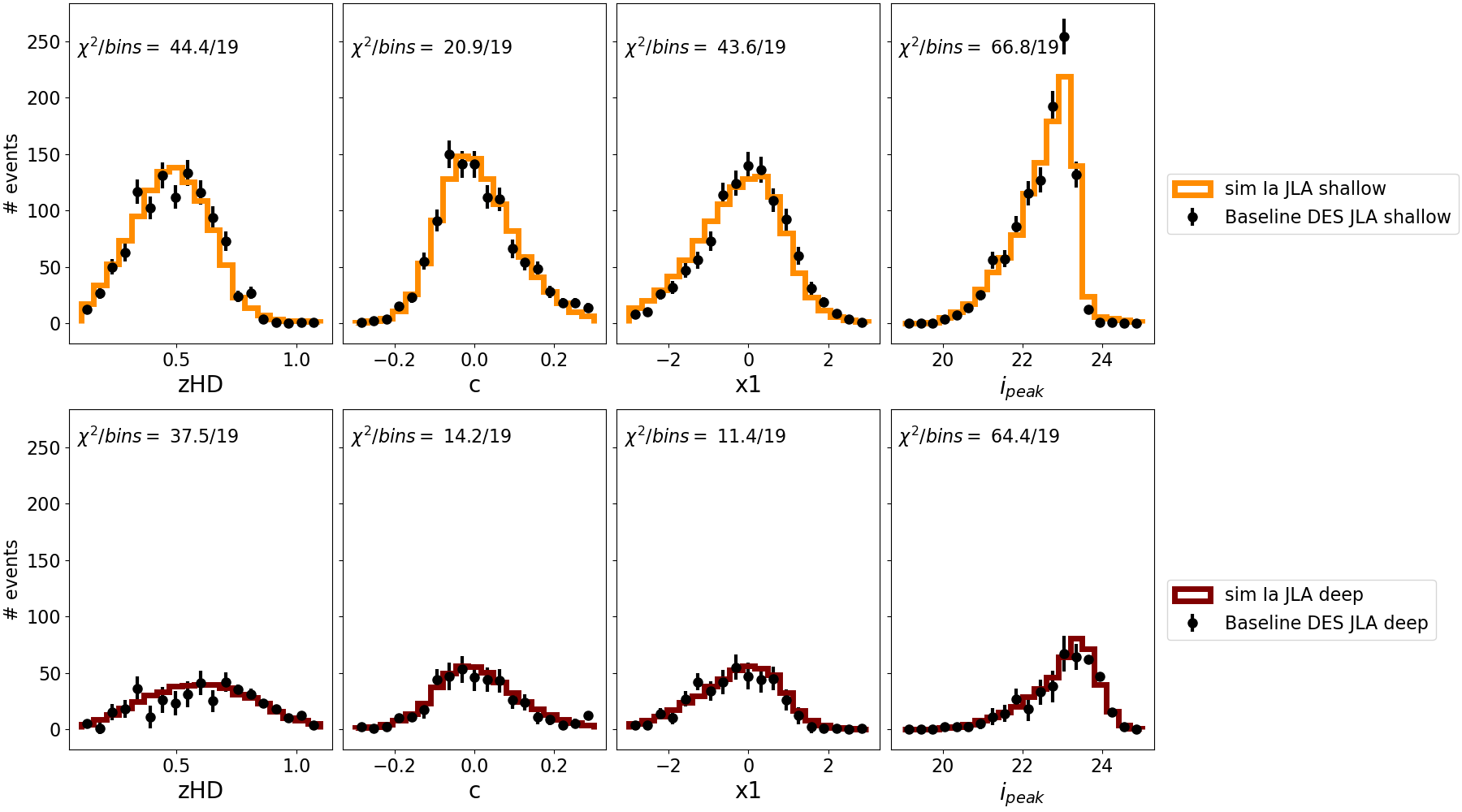}
    \caption{Distributions of redshift, SALT2 x1, SALT2 c and peak magnitude in i-band $i_{\rm peak}$ for our Baseline DES-SNIa sample from Section~\ref{sec:PC_wzspe} for the shallow (yellow) and deep (maroon) fields. We show one simulated realisation of DES-SN 5-year sample. Poisson uncertainties are assumed. Both the simulation and data pass JLA-like cuts. The goodness-of-fit for each histogram is shown as the $\chi^2$/number of bins on each plot.}
    \label{figure:histcx1z}
\end{figure*}

\subsection{Classification efficiency} \label{sec:PCwz_classeff}

Traditionally, in cosmology analyses using spectroscopically classified SNe samples, modelling selection effects is crucial to estimate biases and systematic uncertainties. 

Selection effects arise from a combination of SN detection and other effects. They are usually modelled as an efficiency with respect of an observed magnitude. For host galaxy selection, \cite{Vincenzi:2020} uses the host galaxy $r$ band magnitude, $m^{\rm host}_r$. For spectroscopic classification, \cite{Smith:2020,Kessler:2019} use the modelled supernova peak magnitude in the $i$ band, $i_{\rm peak}$ computed from the best-fit SALT2.

To determine if there is a selection efficiency decrease due to photometric classification, we inspect the differences between the peak observed magnitude in the $i$ band of our Baseline DES-SNIa sample compared to simulated SNe Ia in DES-SN 5-year in Figure~\ref{figure:histcx1z}. Our Baseline DES-SNIa photometric sample follows the expected SN Ia peak magnitude distribution from simulations but we find an excess on the maximum magnitude with a reduced $\chi^2_\nu=2.1$. We do not find evidence for additional selection efficiency effects from the photometric classification procedure.

\subsection{Colour and stretch evolution} \label{sec:PCwz_properties}

We study the properties of the Baseline DES-SNIa sample with JLA-like cuts and compare it to that expected from realistic simulations. In Section~\ref{sec:PCwz_classeff} we found that the effects of classification efficiency are negligible, thus we don't correct for this efficiency and use simulations including only detection and host galaxy redshift efficiency introduced in Section~\ref{sec:sims}.

Figure~\ref{figure:histcx1z} shows the redshift $zHD$ and SALT2 fitted colour $c$, stretch $x_1$ and $i_{\rm peak}$ distributions for the DES-SNIa 5-year photometric sample classified using host galaxy redshifts. Figure~\ref{figure:histcx1z} also shows one realisation of a DES-SN 5-year simulated SNe Ia. Uncertainties are calculated as the square-root of the number of candidates per bin.
There is decent agreement between the simulation and data, although the reduced $\chi^2_\nu$ are somewhat larger than expected from statistical fluctuations.

\begin{figure*}
    \centering
    \includegraphics[width=\textwidth]{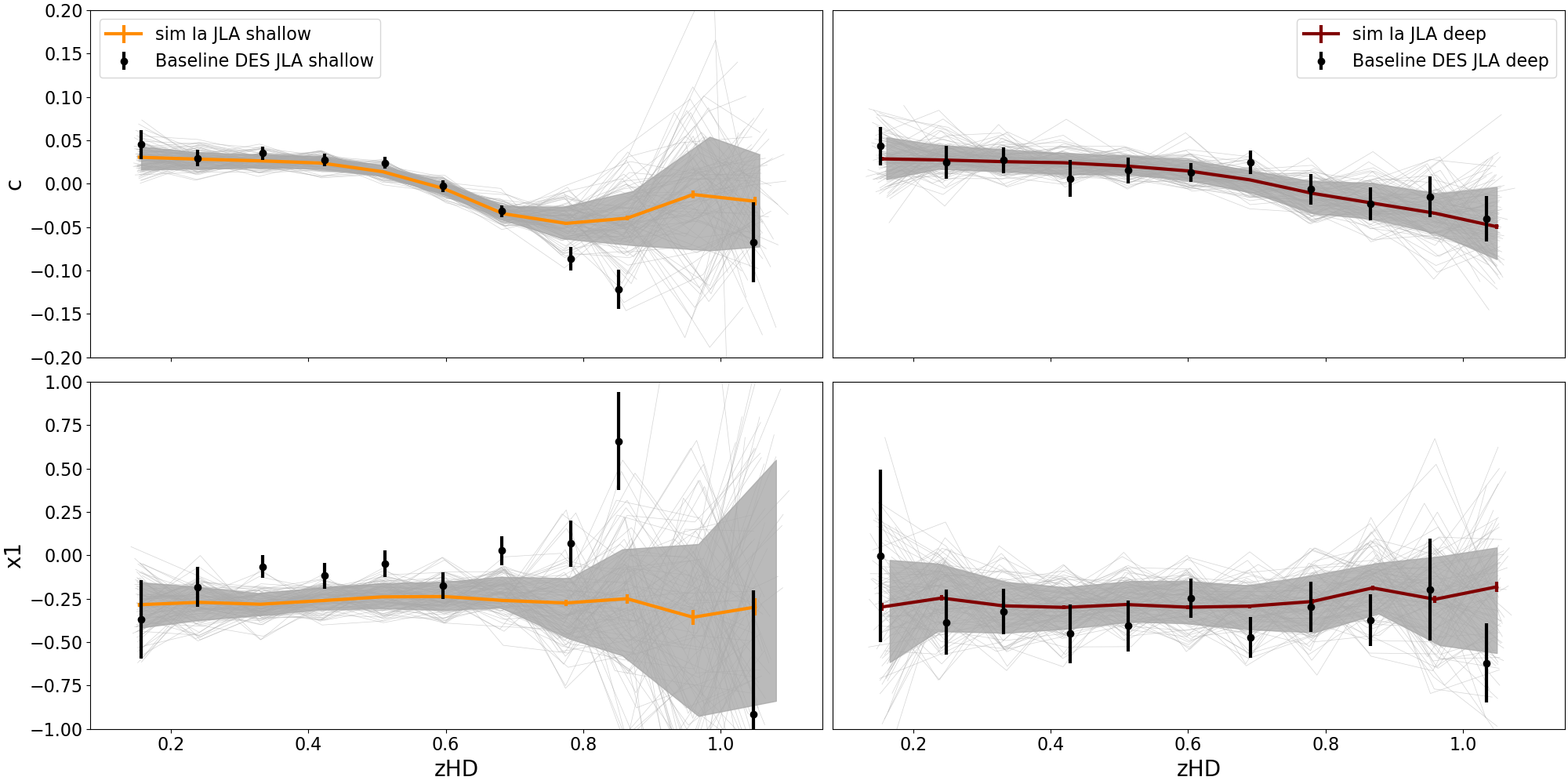}
    \caption{Redshift dependence of SALT2 $c$ and $x_1$ for the Baseline DES-SNIa photometric sample and simulated SNe Ia for shallow (yellow, left) and deep (right, maroon) fields using the DES-SN host galaxy spectroscopic efficiency \citep{Vincenzi:2020} both with JLA-like cuts. For the simulation, orange lines are rolling averages of the measured parameters, in grey 150 realisations of SNe Ia in the DES-SN 5-year survey and in solid grey the area covered by the $68\%$ of these realisations. The mean and the standard deviation are shown for data using black markers.}
    \label{figure:evol_cvsz}
\end{figure*}

In Figure~\ref{figure:evol_cvsz} we show the redshift evolution of our sample's colour and stretch. Our baseline sample matches the trends expected from the simulation. Although there are some slight differences outside the $68\%$ simulation contour (equivalent to $1\sigma$ for a Gaussian distribution) in particular for the shallow fields.

These differences might result from the small number of candidates (the last two redshift bins have only $24$ and $16$ SNe Ia), unaccounted classification contamination, unaccounted selection effects or whether there is redshift evolution in the intrinsic SN population \citep{Scolnic:2016,Popovic:2021,Nicolas:2021} or the effect of dust needs to be introduced \citep{Jha:2007,Mandel:2011,Mandel:2017,Brout:2021}. The optimisation of the simulation and systematics studies is outside the scope of this work.

We now turn to select other photometric samples using the novel Bayesian Neural Networks and explore their possible use.

\section{Photometrically classified SNe Ia with Bayesian Neural Networks}\label{section:PC_BNNs}
In this section we explore the use of Bayesian Neural Networks (BNNs) for classification. While the accuracy of these Networks is equivalent to the baseline RNN used in Section~\ref{sec:PC_wzspe}, BNNs also provide classification uncertainties.

We first obtain photometric samples using two BNN schemes (MC and BBB, Section~\ref{sec:PC_BNN_wz}). We then evaluate the classification uncertainties from BNNs (Section~\ref{sec:PC_BNN_wz_uncertainties}), and summarise our findings (Section~\ref{sec:PC_BNN_wz_summary}).

\subsection{BNN photometric sample}\label{sec:PC_BNN_wz}

We apply our BNN trained models to candidates passing loose and JLA-like cuts introduced Sections~\ref{sec:selcuts_wz} and \ref{sec:JLA-like}. This candidate sample contains 1,701 light-curves that are then photometrically classified.

Using BNN probabilities, the average probability ensemble method and a threshold of $P$ larger than $0.5$, we obtain about $3\%$ more candidates than our Baseline DES-SNIa sample with JLA-like cuts in Table~\ref{tab:PC_wz} for both BNN methods. The additional BNN selected supernovae, 52 MC and 51 BBB, have distributions of colour, stretch and redshifts that are representative of the Baseline DES-SNIa sample selected using the RNN models (Section~\ref{sec:PC_wzspe}). We find that 1 and 6 SNe Ia in the Baseline DES-SNIa sample are not selected by MC and BBB methods. These missing SNe Ia have red colours and are at median redshifts close to $0.5$. The BNN samples are thus probing a similar parameter space to the Baseline DES-SNIa sample.

As in the previous sample, we find no spectroscopically identified AGN, SLSNe or other SN spectral types in our BNN photometric sample. We find the same 5 candidates with nearby spectra showing AGN features which are kept due to their large enough separation $>0.5''$, with the AGN. In a cosmological sample however, these candidates will be eliminated due to possible issues with the measured photometry.

\begin{table}
    \centering
    \caption{Photometric classification of light-curves with Bayesian Neural Networks. Columns indicate: the number of photometrically selected events and the number of spectroscopic SNe Ia contained in that sample. We show these samples with JLA-like SALT2 cuts as in Section~\ref{sec:towards_cosmo} and when adding a cut in the BNN classification uncertainty.}
    \label{tab:PC_wz_BNN}
\begin{tabular}{lllll}

              &     \multicolumn{2}{c}{+JLA-like} &  \multicolumn{2}{c}{+JLA-like +unc} \\
              method &     photo Ia &   spec Ia & photo Ia & spec Ia \\             

\hline
 \multicolumn{5}{c}{MC dropout}\\
 \hline
        single model & $ 1532^{+7}_{-4} $ & $ 335^{+1}_{-1} $ & $        1513^{+6}_{-3} $ & $        333^{+0}_{-0}$ \\
ensemble (prob. av.) & $ 1535^{+3}_{-2} $ & $ 336^{+0}_{-0} $ & $        1520^{+2}_{-1} $ & $        333^{+0}_{-0}$ \\
\\
\textbf{Baseline MC sample} &     1535 &     336 &                 1520 &                 333 \\
\hline
\multicolumn{5}{c}{BBB}\\
\hline

        single model & $ 1526^{+8}_{-6} $ & $ 334^{+1}_{-0} $ & $        1487^{+5}_{-2} $ & $        328^{+2}_{-0}$ \\
ensemble (prob. av.) & $ 1528^{+1}_{-1} $ & $ 335^{+1}_{-0} $ & $        1483^{+0}_{-0} $ & $        324^{+1}_{-0}$ \\ 
\\
\textbf{Baseline BBB sample} &     1529 &     336 &                 1483 &                 324 \\

\end{tabular}
\end{table}

\subsection{BNN uncertainties}\label{sec:PC_BNN_wz_uncertainties}

In this Section we try to interpret which types of uncertainties are captured in the outputs of the BNN model: \textit{aleatoric} or \textit{epistemic}.
BNNs provide classification probability distributions that a priori indicate a confidence level on the prediction. These uncertainties are shown in Figure~\ref{fig:mosaic_lcs_PC} for each classification step. Here we only evaluate the final uncertainty (final time step) for each event. 

\begin{figure}
    \centering
    \includegraphics[width=1.1\columnwidth]{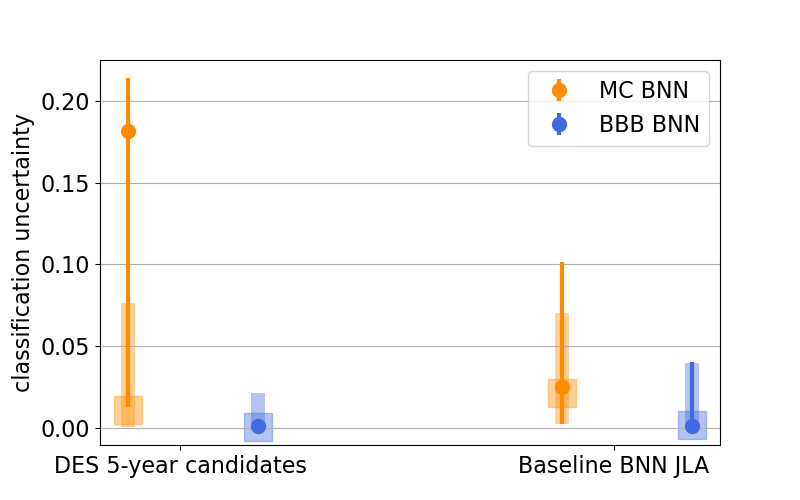}
    \caption{Classification uncertainties obtained for BNN ensemble models. Columns indicate which sample is used. For each event in a given sample, we obtain their classification uncertainties from the two BNN methods, MC and BBB (orange and blue respectively). We show median uncertainties for data in circles for: all DES-SN 5-year data (no selection cuts), and Baseline BNN SNIa samples with JLA-like cuts. For comparison, we show in squares the median uncertainties obtained for the whole simulation (first column) and simulated photometric samples with JLA-like cuts (second column). For both the data and simulations, we show as errorbars the extent of the $68\%$ of the distribution. The different behaviour of simulated MC uncertainties and that of DES-SN 5-year candidate sample is further studied in Figure~\ref{fig:uncertainties}.}
    \label{fig:BNN_uncertainties_allDES}
\end{figure}

In Figure~\ref{fig:BNN_uncertainties_allDES} we show the distribution of classification uncertainties for different samples. We compare the uncertainties derived from the data and from simulations. For most samples, the simulation and data uncertainty distributions are similar. This indicates that the simulations and data resemble closely after JLA-like cuts. However, a large difference is found where there is no selection cut which is further explored in Section~\ref{sec:vro_bnn}. 

Both BNN methods provide different order of magnitude of uncertainties estimates and distribution of mean uncertainties (e.g. BBB is more clustered in low uncertainty regions), possibly due to initialisation parameters or intrinsic properties of the method. Accounting for those differences is not straight-forward, see \cite{Moller:2020} for a discussion on this topic.

\begin{figure*}
    \centering
    \includegraphics[width=\textwidth]{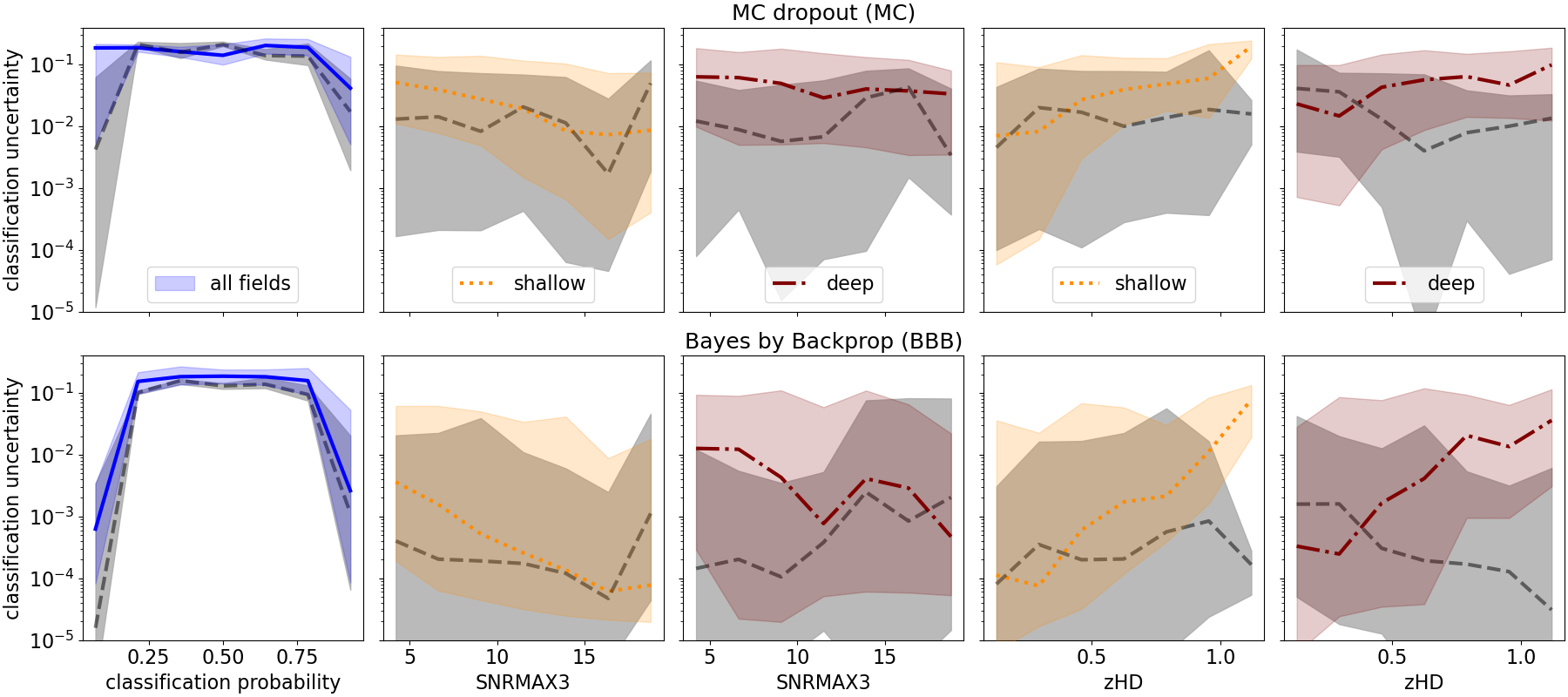}
    \caption{Distribution of classification uncertainty for Baseline MC dropout (upper row) and Bayes by Backprop technique (lower row). We show uncertainties as a function of classification probability for all fields (left), SNR of the third brightest point in the light-curve (SNRMAX3, columns 2 and 3), and redshift (zHD, columns 4 and 5). Coloured lines show the median of the data with solid blue representing all fields, dotted yellow representing shallow fields, and dot-dash red representing the deep fields. Simulations are shown by the grey dashed lines.  Shaded regions show the $68 \%$ percentile. }
    \label{fig:BNN_uncertainties}
\end{figure*}

We compare BNN uncertainties as a function of light-curves properties in Figure~\ref{fig:BNN_uncertainties}.  We find that MC dropout and BBB exhibit different behaviours for both data and simulations.

We find both indications in favour (+) and against (-) interpretation of classification uncertainties as a particular type:
\begin{itemize}
    \item[a.] \textit{aleatoric uncertainty}: linked to measurement uncertainties\\
    (+) classification uncertainties are correlated to SNR in data. Bright candidates and those with higher quality light-curves have on average smaller classification uncertainties for both BNNs.\\
    (-) this correlation is not seen in the simulations for any of the BNNs.\\
    \item[b.] \textit{epistemic uncertainty}: linked to training sets or model \\
    (+) Large uncertainties are more prevalent in classification probabilities far from 1 (high probability of being a SN Ia) and 0 (low probability of being SN Ia) for both simulations and DES-SN 5-year data. \\
    (-) candidates that fulfil selection cuts should more closely resemble simulated SNe Ia, thus it is puzzling the increase on median uncertainty when applying cuts in particular for the MC method (see Figure~\ref{fig:BNN_uncertainties_allDES}).
\end{itemize}
These various behaviours highlights the challenges on quantifying uncertainties in complex problems such as astronomical data classification. In Appendix~\ref{appendix:uncertainties} we explore further correlations between classification uncertainties and SALT2 fit light-curve properties.

We continue exploring the interpretability of the BNNs uncertainties by adding a threshold on the uncertainties for SNIa sample selection, as in \cite{Moller:2020} and more recently in \cite{Butter:2021}. We note that establishing a threshold for uncertainties is not straight-forward. While the whole probability distribution has a calibration that can be verified using diagnostic as reliability diagrams \citep{DeGroot:1983,Moller:2020}, the probability uncertainties do not. We chose to eliminate candidates with the highest uncertainties (eliminating candidates that are outside of 99 percentile of the uncertainty distribution). This cut rejects candidates that were in the RNN sample: 12 for the MC model and 45 for BBB. These candidates are not found to be distributed preferentially in a $c$, $x_1$ or redshift. We visually inspect these light-curves and found that a large proportion have photometry that are outliers.

\subsection{BNN photometric sample contribution} \label{sec:PC_BNN_wz_summary}

The SNIa samples obtained using BNN methods are found to be similar to the one provided by our Baseline DES-SNIa sample in Section~\ref{sec:PC_wzspe}. We evaluate BNN uncertainties and show that they are consistent between simulations and data in average after JLA-like cuts, showing a good agreement between data and simulation predictions. However, BNN uncertainties are difficult to interpret and assess quantitatively (e.g. assigning an uncertainty threshold).

We find that uncertainties exhibit different behaviours in the two BNN methods and between data and simulations. While the higher uncertainties in the MC BNN method for the data could point towards the presence of out-of-distribution candidates, the evidence is not conclusive and is not seen in the BBB method. We will further explore the possible contribution of BNNs in photometric classification without any selection cuts in Section~\ref{sec:vro_bnn}.

Cuts on uncertainty values potentially improve our photometric SNIa samples by rejecting candidates with photometry that contains outliers. These is a promising avenue shown to improve the quality of samples, both in quality of the data and rejection of out-of-distribution events, in previous work using simulations \cite{Moller:2020} and more recently with astronomical data in \cite{Butter:2021}.

\section{From DES to Rubin Observatory LSST} \label{sec:lsst}
For the LSST survey, where up $10^7$ SNe will be detected over 10 years, photometric classification will  become increasingly important.

In this work, we have presented different methods for photometric classification with redshift information. We compare the samples obtained with these different methods in Section~\ref{sec:vro_z_spec} and explore possible applications of Bayesian Neural Networks in future surveys, such as LSST, in Section~\ref{sec:vro_bnn}.

\subsection{DES-SNIa photometric samples}\label{sec:vro_z_spec}

The DES-SN 5-year data contains thousands of potential SNe Ia. We show in Table~\ref{tab:summary_selcuts_PC} the different steps used in this work to obtain our Baseline DES-SNIa JLA sample from the DES-SN 5-year candidate sample. Cuts applied before photometric classification reduce the candidate sample by 90\%. Photometric classification and JLA-like cuts refine the sample with a small 20\% reduction. While this reduction is small, it reduces contamination from $\sim 10\%$ to below 1.4\%, as shown in \citep{Vincenzi:2021} and in Section~\ref{table:metrics_wcuts}.

\begin{figure*}
    \centering
    \includegraphics[width=\textwidth]{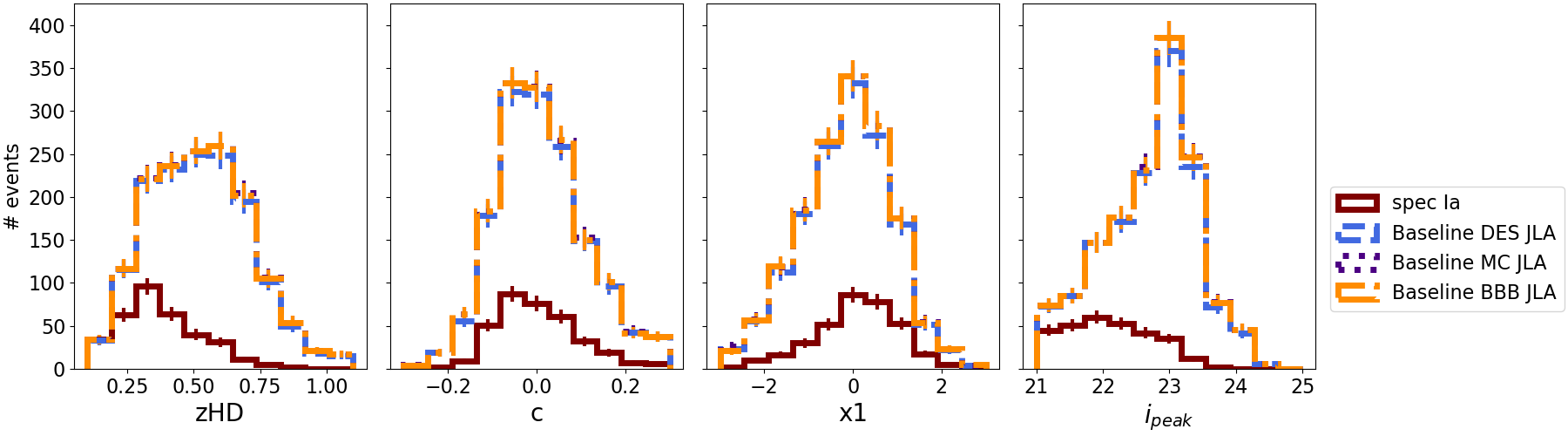}
    \caption{Distributions of redshift, SALT2 c, x1 and peak magnitude in the $i$ band $i_{\rm peak}$, for the samples with JLA cuts: preliminary DES-SN 5-year SNIa spectroscopically classified (maroon), Baseline DES-SNIa (RNN) (blue) and Baseline MC and BBB SNIa samples, purple and orange, respectively. We note that the MC and BBB samples distributions almost completely overlap one other.}
    \label{fig:hist_samples}
\end{figure*}

In addition to our Baseline DES-SNIa sample classified using RNN probabilities, we have explored identifying samples with Bayesian Neural Networks. We compare these samples with with the preliminary DES-SN 5-year spectroscopically classified SNe Ia sample in Figure~\ref{fig:hist_samples}. As expected, we find that photometric samples using RNNs or BNNs provide larger numbers of SNe Ia than the spectroscopic sample, probing a larger parameter space. We do not find a substantial difference in the parameter distributions between different photometric classification methods. 

We highlight that the photometric samples peak at fainter magnitudes and higher redshifts than the preliminary DES-SN 5-year spectroscopic SNe Ia sample.This has the potential to reduce selection biases and opens the possibility of stronger statistical analyses with the large numbers of SNe Ia. This will also be true for the immense SN samples obtained with LSST.

\subsection{Bayesian Neural Networks as a proxy}\label{sec:vro_bnn}

Introduced as a promising method to quantify model uncertainties, BNNs have not yet been widely used in classification tasks. In Section~\ref{section:PC_BNNs}, we have shown the difficulties for uncertainty interpretation given the different uncertainty values for the BNN methods. However, a potential use could be rejecting candidates with large uncertainties, as they sometimes have light-curves with photometry outliers.

Here, we explore other possible uses of BNN uncertainties, using samples that have not been constrained with selection cuts. We aim to answer two questions: (i) can BNN uncertainties be used as an indicator of the representativity of the training set for a given dataset?  (ii) can BNN uncertainties replace selection cuts? We address these questions in Sections~\ref{sec:BNN_representativity} and \ref{sec:bnn_cuts} respectively. The former could be useful to choose the set of SED templates to simulate a survey. As some selection cuts require feature extraction, the latter could be valuable to avoid this time-consuming process by using instead classification uncertainties from non-parametric classifiers as \ssnn.

\subsubsection{BNNs uncertainties vs. simulation representativity} \label{sec:BNN_representativity}

First, we use simulations to assess the expected behaviour of uncertainties when training sets are not representative of the testing data. 

We examine how the uncertainties change when using the trained model in Section~\ref{sec:BNN_probs} and applied to individual simulations with normal Type Ia supernovae and core-collapse SNe generated with the V19, SPCC and J17 templates. We expect that the trained model is representative of the V19 simulation. This will not be true for J17 and SPCC.

We find that both the single seed and ensemble methods have accuracies which decrease for J17 and SPCC simulations by $\approx 0.5\%$ for both types of BNNs. We see an increase in the {\it mean model uncertainty} on classified light-curves generated with J17 and SPCC, however this change is within uncertainties. For both BNNs we find a longer and more significant tail for the uncertainty distributions when classifying J17 and SPCC simulations (ending at $\sim0.4-0.43$ compared to $\sim0.35$ for V19).

Next, we compare uncertainties when classifying DES-SN 5-year data with independent BNN models trained with the V19, J17 and SPCC simulations. We find that the {\it mean model uncertainty} increases for SPCC and J17 classification models for MC dropout but not for BBB SPCC model but again within uncertainties. The tail of the uncertainties varies between $\sim 0.40-0.47$ for all classification models. We see a longer tail for the uncertainty distributions for BBB but not for MC SPCC classification.

In summary, we do not find strong evidence of BNN uncertainties being sensitive to models trained with different core-collapse templates. There is a small but inconclusive tendency to increase uncertainties for J17 and SPCC in simulations. While these templates are different, the changes may be too small to be captured by BNN uncertainties.

\subsubsection{BNN uncertainties as a proxy for selection cuts?}\label{sec:bnn_cuts}

We further study the distribution of classification uncertainties for samples selected with different cuts.

First, we check the behaviour of uncertainties with simulations. Uncertainties are distributed with a peak at low values and a decreasing long tail. We find that as the sample is refined through cuts in redshift, SALT2 convergence, and others, the maximum uncertainty is reduced. For example, if the simulated sample passes loose selection cuts and then a JLA-like cut is applied, the maximum uncertainty in the distribution reduces from $0.37$ to $0.26$ in MC dropout and from $0.34$ to $0.25$ in BBB. We do not find a significant change in the median distribution since it is dominated by small uncertainty values.

\begin{figure*}
    \centering
    \includegraphics[width=\textwidth]{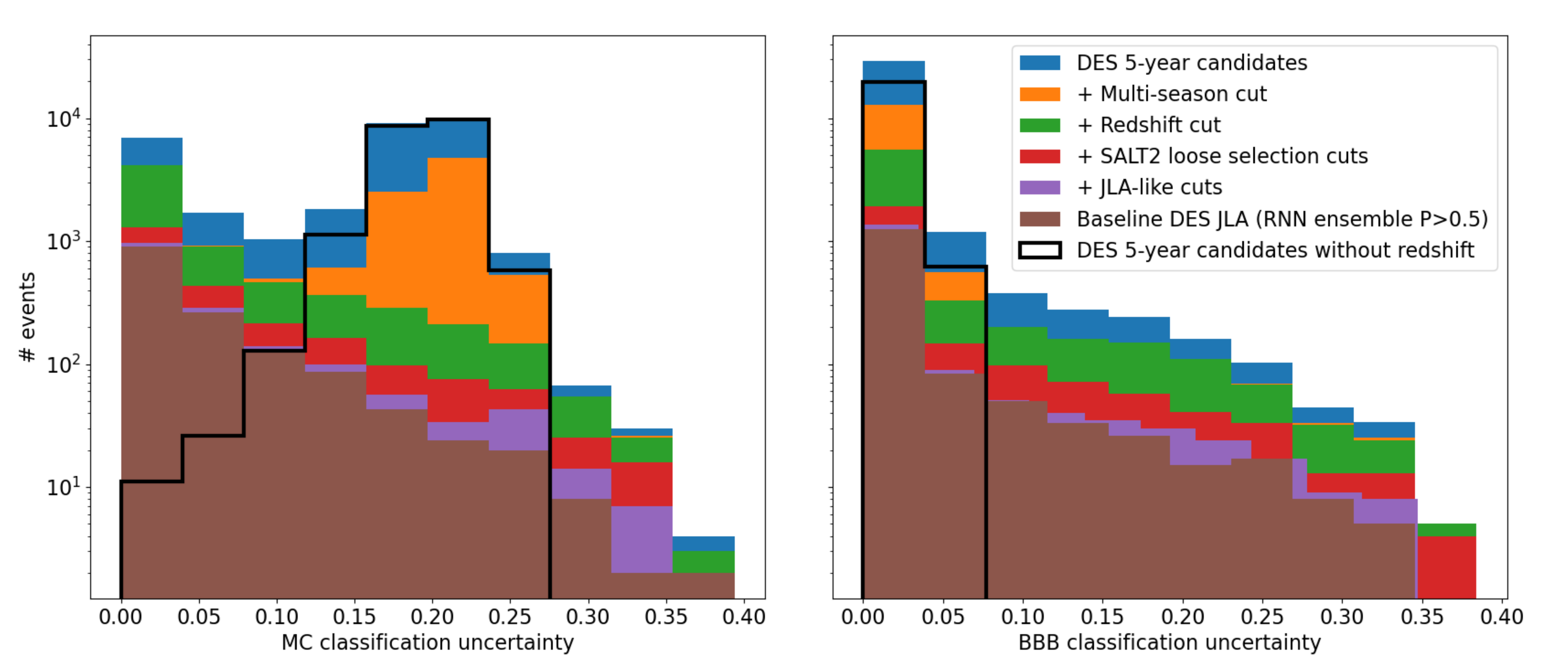}
    \caption{Uncertainties obtained with the two BNN methods (MC and BBB) for the DES-SN 5-year candidate sample, through different selection cuts: multi-season filtering, redshifts, SALT2 convergence and JLA-like cuts; and our photometrically identified sample (filled histograms). We show the number of events in the y-axis in log scale. MC dropout uncertainties seem to identify those out-of-distribution candidates that have no redshift information (black line) or are filtered multi-season events. This secondary peak drives the mean uncertainty behaviour for MC dropout in Figure~\ref{fig:BNN_uncertainties_allDES}.}
    \label{fig:uncertainties}
\end{figure*}

For the DES-SN 5-year data we show the distribution of classification uncertainties in Figure~\ref{fig:uncertainties} with different selection cuts (see Section~\ref{sec:selcuts_wz}). As selection cuts are applied, the maximum uncertainties reduces for both methods as in simulations.

We highlight an interesting behaviour seen for MC dropout classification uncertainties. We find that this method assigns high uncertainties to candidates that do not have a secured redshift and candidates that are filtered with the multi-season cut. While the model was trained to use host galaxy redshifts, it can provide a classification for objects using a default value provided, here an assigned redshift of $-9$. While these candidates are clearly outliers (the redshift provided for classification is -9) and can be eliminated using simple cuts, this could indicate that MC dropout uncertainties are indicative of out-of-distribution candidates. Importantly, many of these high-uncertainty candidates are classified with probability larger than $0.5$ which, without selection cuts, would end up in our photometric sample if no selection cuts were applied. We do not see this behaviour in the BBB model. 

The multi-season veto and redshift availability cut effectively eliminates the light-curves producing the high-uncertainty peak for MC dropout. After these cuts, the most impactful cut for higher uncertainties is linked to the SALT and JLA quality cuts. This is not surprising since these cuts restrict the SN properties range to the ones for normal SNe Ia.

In summary, we find that BNN methods behave differently when classifying out-of-distribution candidates defined as light-curves without redshift. Interestingly, the high-uncertainty peak found for the MC dropout method in Figure~\ref{fig:uncertainties} reflects a possible interpretability of these uncertainties. This interpretability could help to quickly identify the presence of anomalies in the dataset which were not in the training sets of the model. 

For current surveys, our candidate samples are small enough to easily identify out-of-distribution events using feature distributions. However, for future surveys such as Rubin LSST this may prove difficult given the expected detection of 10 million transient candidates per night. Here we find that BNN uncertainties from MC dropout scheme can provide an indication whether there are out-of-distribution events in a given candidate sample and further selection cuts may be required.

\section{Conclusions}

In this work we train Type Ia vs. non Ia classification models using large realistic DES-like simulations and apply them to DES-SN 5-year data.

We introduce pre-processing of DES-SN light-curves for accurate photometric classification. This includes selection of light-curve time-span, photometry quality cuts and selection cuts to limit out-of-distribution candidates that are not included in the training set (e.g. AGNs).

We present samples classified with host galaxy redshifts using \ssnn Recurrent Neural Networks and explore the use of Bayesian Neural Networks. We introduce the use of ensemble predictions for SN classification. We find that selecting SNe using an ensemble of models is more robust and stable than any single model.

Using host galaxy spectroscopic redshifts, we select a Baseline DES-SNIa sample of 1,863 photometrically identified Type Ia SNe. This sample can be used for astrophysical studies of the properties of SNe Ia and their environments. For cosmology, we apply JLA-like cuts and select $1,484$ photometrically classified SNe Ia. This sample is more than three times larger than the DES-SN 5-year spectroscopically confirmed SN Ia sample and covers a larger redshift range. Most of the spectroscopically identified SNe Ia in DES-SN are included in this photometric sample. These 1,484 photometrically identified SNe Ia are currently the largest single-survey high-quality SN Ia sample and is being used for studies such as rates and SNe Ia host-galaxy properties.

We find that the properties of the SNe Ia in our Baseline DES-Ia sample are reproduced in the simulations. We anticipate that with further refinements (improved host galaxy libraries and more accurate dust models), the agreement between the simulations and the data will improve.

Additionally, we explore the use of uncertainties provided by Bayesian Neural Networks for identifying out-of-distribution candidates and defining representative training sets. We highlight some of the BNN pitfalls and the difficulty of comparing classification uncertainties between variational inference methods. We find that the MC dropout BNN provides potentially interpretable uncertainties for out-of-distribution event detection and improving the photometric sample. This work is the first known application of two BNN methods on real astrophysical data for classification tasks. 

This work is part of the DES-SN 5-year cosmology analysis. We have optimised simulations, the \ssnn architecture, as well as developed data pre-processing methods. These methods are a revision from those presented in \cite{Vincenzi:2021} where contamination is found to be less than 1.4\% for photometrically classified samples. We find that photometric quality is key for robust classification, and an improved sample can be expected from using high-quality Scene Modelling Photometry \citep{Brout:2019}.

For future surveys such as LSST, photometric classification will be key to fully harness the power of these surveys. Photometric classification with host redshift information will enable using large, low-contamination, high-quality samples for measuring cosmological parameters. Potentially, MC BNN could provide useful information to filter transient samples in large surveys. Extensions to this work include photometric classification without redshift, which will assist in the allocation of follow-up resources for host galaxy redshift acquisition \citep[such as Time-Domain Extragalactic Survey TiDES; Frohmaier et al. in prep,][]{Swann:2019} and for other astrophysical studies.

\section*{Acknowledgements}

AM thanks of the University of Queensland for hospitality during early work on this project; parts of this research were supported by the Australian Research Council Australian Laureate Fellowship FL180100168. 

This paper has gone through internal review by the DES collaboration. Funding for the DES Projects has been provided by the U.S. Department of Energy, the U.S. National Science Foundation, the Ministry of Science and Education of Spain, the Science and Technology Facilities Council of the United Kingdom, the Higher Education Funding Council for England, the National Center for Supercomputing Applications at the University of Illinois at Urbana-Champaign, the Kavli Institute of Cosmological Physics at the University of Chicago, the Center for Cosmology and Astro-Particle Physics at the Ohio State University, the Mitchell Institute for Fundamental Physics and Astronomy at Texas A\&M University, Financiadora de Estudos e Projetos, Fundação Carlos Chagas Filho de Amparo à Pesquisa do Estado do Rio de Janeiro, Conselho Nacional de Desenvolvimento Científico e Tecnológico and the Ministério da Ciência, Tecnologia e Inovação, the Deutsche Forschungsgemeinschaft and the Collaborating Institutions in the Dark Energy Survey.
The Collaborating Institutions are Argonne National Laboratory, the University of California at Santa Cruz, the University of Cambridge, Centro de Investigaciones Energéticas, Medioambientales y Tecnológicas-Madrid, the University of Chicago, University College London, the DES-Brazil Consortium, the University of Edinburgh, the Eidgenössische Technische Hochschule (ETH) Zürich, Fermi National Accelerator Laboratory, the University of Illinois at Urbana-Champaign, the Institut de Ciències de l’Espai (IEEC/CSIC), the Institut de Física d’Altes Energies, Lawrence Berkeley National Laboratory, the Ludwig-Maximilians Universität München and the associated Excellence Cluster Universe, the University of Michigan, NFS’s NOIRLab, the University of Nottingham, The Ohio State University, the University of Pennsylvania, the University of Portsmouth, SLAC National Accelerator Laboratory, Stanford University, the University of Sussex, Texas A\&M University, and the OzDES Membership Consortium.

Based in part on observations at Cerro Tololo Inter-American Observatory at NSF’s NOIRLab (NOIRLab Prop. ID 2012B-0001; PI: J. Frieman), which is managed by the Association of Universities for Research in Astronomy (AURA) under a cooperative agreement with the National Science Foundation.

The DES data management system is supported by the National Science Foundation under Grant Numbers AST-1138766 and AST-1536171. The DES participants from Spanish institutions are partially supported by MICINN under grants ESP2017-89838, PGC2018-094773, PGC2018-102021, SEV-2016-0588, SEV-2016-0597, and MDM-2015-0509, some of which include ERDF funds from the European Union. IFAE is partially funded by the CERCA program of the Generalitat de Catalunya. Research leading to these results has received funding from the European Research Council under the European Union’s Seventh Framework Program (FP7/2007-2013) including ERC grant agreements 240672, 291329, and 306478. We acknowledge support from the Brazilian Instituto Nacional de Ciência e Tecnologia (INCT) do e-Universo (CNPq grant 465376/2014- 2).

This work was completed in part with Midway resources provided by the University of Chicago’s Research Computing Center.

This work makes use of data acquired at the Anglo-Australian Telescope, under program A/2013B/012. We acknowledge the traditional owners of the land on which the AAT stands, the Gamilaraay people, and pay our respects to elders past and present.

MS is funded by the European Reearch Council (ERC) under the European Union's Horizon 2020 Research and Innovation program (grant agreement no 759194 - USNAC). L.G. acknowledges financial support from the Spanish Ministerio de Ciencia e Innovaci\'on (MCIN), the Agencia Estatal de Investigaci\'on (AEI) 10.13039/501100011033, and the European Social Fund (ESF) "Investing in your future" under the 2019 Ram\'on y Cajal program RYC2019-027683-I and the PID2020-115253GA-I00 HOSTFLOWS project, and from Centro Superior de Investigaciones Cient\'ificas (CSIC) under the PIE project 20215AT016. LK thanks the UKRI Future Leaders Fellowship for support through the grant MR/T01881X/1. 

\section*{Data Availability}

We provide in \url{https://github.com/anaismoller/DES5YR_SNeIa_hostz}: (i) the SNANA and/or Pippin configuration files to reproduce simulations in this paper, (ii) configuration files and scripts to re-train \ssnn classifications models (SNN is an open source framework available in GitHub), and (iii) analysis code in python to reproduce plots and results. Sample classification probabilities are available in Zenodo \url{https://doi.org/10.5281/zenodo.5904368}.




\bibliographystyle{mnras}
\bibliography{ref} 




\appendix

\section{Uncertainties and fitted parameters}\label{appendix:uncertainties}
\begin{figure*}
    \centering
    \includegraphics[width=\textwidth]{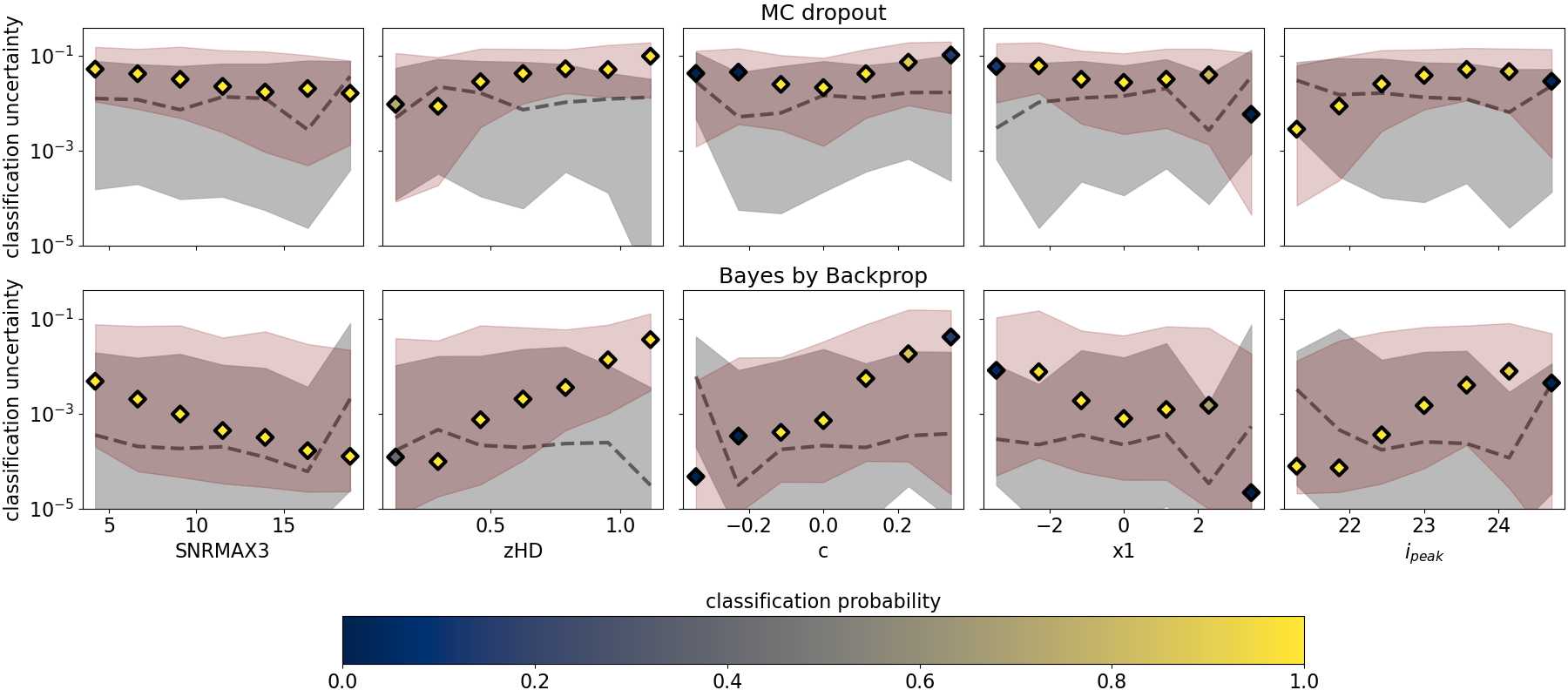}
    \caption{Distribution of classification uncertainties for DES 5-year data (maroon) and simulation (grey) using the two BNNs, MC dropout and Bayes by Backprop. We show uncertainties as a function of SNR of the third brightest point in the light-curve, redshift, colour, stretch and peak $i$-band magnitude. The median and $68$ percentile are shown as a dashed line and filled coloured area. 
    Data is shown as diamonds coloured by the median probability in that parameter bin and a maroon coloured area representing the $68\%$ percentile of the distribution.}
    \label{fig:BNN_uncertainties_SALT2}
\end{figure*}

In Section~\ref{sec:PC_BNN_wz_uncertainties} we explored the interpretability of BNN uncertainties. We concluded that this interpretation was not straight forward from our results. Here we extend this discussion by exploring possible correlations with other light-curve properties derived from a SALT2 fit in Figure~\ref{fig:BNN_uncertainties_SALT2}.

In general, we find that uncertainties tend to be larger for the data when compared with simulations. The uncertainties in BBB method varies more with the parameters. 

We note that the classification uncertainties are large for red and high stretch SNe in the DES 5-year sample. The median classification probability is also lower for these candidates. If the uncertainties are epistemic due to a smaller training set, then they would be large for the ends of the normal SNe Ia SALT2 parameter distributions since training sets have fewer such candidates. However, we do not find this behaviour. Another possible effect could be that bluer SNe Ia are more easily standardisable as previous literature suggests and thus their classification is more robust \citep{Kelsey:2021,Brout:2021}. However, as this tendency is only observed in data and not simulations, no conclusion can be confidently drawn.

The peak magnitude in i-band behaviour in data agrees with that of the SNR of the light-curve. Brighter candidates are classified with higher confidence than fainter ones. However, as in the previous Section we do not see such a behaviour in the simulation.

While the correlation between supernova properties and classification uncertainties are interesting to explore, they are difficult to interpret since multiple effects could be contributing to the uncertainties. Tests based on simple physical systems could provide hints towards further interpretability, such as recent work by \cite{Caldeira:2020}.

\section{DES 5-year photometrically selected SNe Ia}\label{appendix:table_probs}

A Table with photometrically classified SNe Ia from all selection methods with their respective probabilities for a subsample of DES 5-year data is provided at \url{https://doi.org/10.5281/zenodo.5904368}. Samples are selected using $P$ larger than $0.5$ for each method plus selection cuts.


\begin{table*}
 \caption{Example of \ssnn classification probabilities for DES 5-year candidates. A full list of classification probabilities for all DES 5-year candidates that are selected for any of the samples in this paper is found in \url{https://doi.org/10.5281/zenodo.5904368}.
 We show probabilities for baseline Recurrent Neural Network (RNN) in Section~\ref{sec:PC_wzspe}) and Bayesian Neural Networks methods MC dropout (MC) and Bayes by Backprop (BBB) in Section~\ref{section:PC_BNNs}. For each method we provide classification probabilities rounded in two decimals for five different \ssnn initialisation seeds, $s_0=0$, $s_1=55$, $s_2=100$, $s_3=1000$, $s_4=30469$ and the ensemble average probability of these five seeds $set0$.}
  \label{tab:PIa}
  
\begin{tabular}{l|llllll|llllll|llllll}

    IAUC & \multicolumn{6}{c}{RNN} & \multicolumn{6}{c}{MC} & \multicolumn{6}{c}{BBB}\\
          &  $s_0$ &  $s_1$ &  $s_2$ &  $s_3$ & $s_4$ &  $set0$ &   $s_0$ &  $s_1$ &  $s_2$ &  $s_3$ & $s_4$ &  $set0$ & $s_0$ &  $s_1$ &  $s_2$ &  $s_3$ & $s_4$ &  $set0$ \\ 
\hline
DES17S2gpk       &       0.99 &        0.95 &         0.99 &          0.99 &           1.00 &          0.98 &      1.00 &       0.96 &        0.98 &         0.98 &          0.99 &         0.98 &       0.97 &        1.00 &         0.92 &          0.99 &           0.98 &          0.97 \\
DES14S2bck       &       1.00 &        1.00 &         1.00 &          1.00 &           1.00 &          1.00 &      1.00 &       1.00 &        1.00 &         1.00 &          1.00 &         1.00 &       1.00 &        1.00 &         1.00 &          1.00 &           1.00 &          1.00 \\
DES14S2anv       &       1.00 &        1.00 &         1.00 &          1.00 &           1.00 &          1.00 &      1.00 &       1.00 &        1.00 &         1.00 &          1.00 &         1.00 &       1.00 &        1.00 &         1.00 &          1.00 &           1.00 &          1.00 \\
DES15S2mji       &       1.00 &        1.00 &         1.00 &          1.00 &           1.00 &          1.00 &      1.00 &       1.00 &        1.00 &         1.00 &          1.00 &         1.00 &       1.00 &        1.00 &         1.00 &          1.00 &           1.00 &          1.00 \\
DES13X3woy       &       1.00 &        1.00 &         1.00 &          1.00 &           1.00 &          1.00 &      1.00 &       1.00 &        1.00 &         1.00 &          1.00 &         1.00 &       1.00 &        1.00 &         1.00 &          1.00 &           1.00 &          1.00 \\
DES14S2aoi       &       1.00 &        1.00 &         1.00 &          1.00 &           1.00 &          1.00 &      1.00 &       1.00 &        1.00 &         1.00 &          1.00 &         1.00 &       1.00 &        1.00 &         1.00 &          1.00 &           1.00 &          1.00 \\
DES13C3xhy       &       0.73 &        0.99 &         1.00 &          0.32 &           0.00 &          0.61 &      0.82 &       0.99 &        0.54 &         0.94 &          0.60 &         0.78 &       0.46 &        0.82 &         0.85 &          0.98 &           0.63 &          0.75 \\
DES15X3dyt       &       1.00 &        1.00 &         1.00 &          1.00 &           1.00 &          1.00 &      1.00 &       1.00 &        1.00 &         1.00 &          1.00 &         1.00 &       1.00 &        1.00 &         1.00 &          1.00 &           1.00 &          1.00 \\
DES14E1gvc       &       1.00 &        1.00 &         1.00 &          1.00 &           1.00 &          1.00 &      1.00 &       1.00 &        1.00 &         1.00 &          1.00 &         1.00 &       1.00 &        1.00 &         1.00 &          1.00 &           1.00 &          1.00 \\
DES13S2ead       &       0.99 &        1.00 &         1.00 &          1.00 &           0.98 &          0.99 &      1.00 &       0.98 &        1.00 &         1.00 &          0.99 &         0.99 &       0.99 &        1.00 &         0.98 &          0.98 &           0.98 &          0.99 \\
DES16S1byw       &       1.00 &        1.00 &         1.00 &          1.00 &           1.00 &          1.00 &      1.00 &       1.00 &        1.00 &         1.00 &          1.00 &         1.00 &       1.00 &        1.00 &         1.00 &          1.00 &           1.00 &          1.00 \\
DES16E2bp        &       1.00 &        1.00 &         1.00 &          1.00 &           1.00 &          1.00 &      1.00 &       1.00 &        1.00 &         1.00 &          1.00 &         1.00 &       1.00 &        1.00 &         1.00 &          1.00 &           1.00 &          1.00 \\
DES15E1nzd       &       1.00 &        1.00 &         1.00 &          1.00 &           1.00 &          1.00 &      1.00 &       1.00 &        1.00 &         1.00 &          1.00 &         1.00 &       1.00 &        1.00 &         1.00 &          1.00 &           1.00 &          1.00 \\
DES16X3enk       &       1.00 &        1.00 &         1.00 &          1.00 &           1.00 &          1.00 &      1.00 &       1.00 &        1.00 &         1.00 &          1.00 &         1.00 &       1.00 &        1.00 &         1.00 &          1.00 &           1.00 &          1.00 \\
DES16X3hy        &       1.00 &        1.00 &         1.00 &          1.00 &           1.00 &          1.00 &      1.00 &       1.00 &        1.00 &         1.00 &          1.00 &         1.00 &       1.00 &        1.00 &         1.00 &          1.00 &           1.00 &          1.00 \\
DES16X3hi        &       1.00 &        1.00 &         1.00 &          1.00 &           1.00 &          1.00 &      1.00 &       1.00 &        1.00 &         1.00 &          1.00 &         1.00 &       1.00 &        1.00 &         1.00 &          1.00 &           1.00 &          1.00 \\
DES15C2mcu       &       1.00 &        1.00 &         1.00 &          1.00 &           1.00 &          1.00 &      1.00 &       1.00 &        1.00 &         1.00 &          1.00 &         1.00 &       1.00 &        1.00 &         1.00 &          1.00 &           1.00 &          1.00 \\
DES13X1hxs       &       1.00 &        1.00 &         1.00 &          1.00 &           1.00 &          1.00 &      1.00 &       1.00 &        1.00 &         1.00 &          1.00 &         1.00 &       1.00 &        1.00 &         1.00 &          1.00 &           1.00 &          1.00 \\
DES13X1bama      &       0.99 &        0.98 &         0.99 &          0.99 &           1.00 &          0.99 &      0.95 &       0.94 &        0.92 &         0.96 &          0.94 &         0.94 &       0.88 &        0.88 &         0.66 &          0.93 &           0.89 &          0.85 \\
DES17C2acb       &       1.00 &        1.00 &         1.00 &          1.00 &           1.00 &          1.00 &      1.00 &       1.00 &        1.00 &         1.00 &          1.00 &         1.00 &       1.00 &        1.00 &         1.00 &          1.00 &           1.00 &          1.00 \\
DES16C3bab       &       1.00 &        1.00 &         1.00 &          1.00 &           1.00 &          1.00 &      1.00 &       1.00 &        1.00 &         1.00 &          1.00 &         1.00 &       1.00 &        1.00 &         1.00 &          1.00 &           1.00 &          1.00 \\
DES17E2elx       &       1.00 &        1.00 &         1.00 &          1.00 &           1.00 &          1.00 &      1.00 &       1.00 &        1.00 &         1.00 &          1.00 &         1.00 &       1.00 &        1.00 &         1.00 &          1.00 &           1.00 &          1.00 \\
DES14E2fyd       &       1.00 &        1.00 &         1.00 &          1.00 &           1.00 &          1.00 &      1.00 &       1.00 &        1.00 &         1.00 &          1.00 &         1.00 &       1.00 &        1.00 &         1.00 &          1.00 &           1.00 &          1.00 \\
DES16C3bq        &       0.97 &        0.99 &         0.79 &          0.31 &           1.00 &          0.81 &      0.95 &       0.83 &        0.91 &         0.94 &          1.00 &         0.93 &       0.78 &        0.95 &         0.69 &          0.94 &           0.87 &          0.85 \\
DES15C3mes       &       1.00 &        1.00 &         1.00 &          1.00 &           1.00 &          1.00 &      1.00 &       1.00 &        1.00 &         1.00 &          1.00 &         1.00 &       1.00 &        1.00 &         1.00 &          1.00 &           1.00 &          1.00 \\
DES15C3meu       &       1.00 &        0.99 &         1.00 &          1.00 &           0.98 &          0.99 &      1.00 &       0.97 &        0.90 &         0.89 &          0.99 &         0.95 &       1.00 &        0.98 &         0.89 &          1.00 &           0.98 &          0.97 \\
DES16X3brw       &       0.79 &        0.02 &         0.35 &          0.02 &           0.25 &          0.29 &      0.99 &       0.94 &        0.96 &         0.97 &          0.71 &         0.92 &       0.91 &        0.95 &         0.64 &          0.86 &           0.73 &          0.82 \\
DES14X1tbo       &       1.00 &        1.00 &         1.00 &          1.00 &           1.00 &          1.00 &      1.00 &       1.00 &        1.00 &         1.00 &          1.00 &         1.00 &       0.99 &        1.00 &         1.00 &          1.00 &           1.00 &          1.00 \\
DES15X3dyv       &       1.00 &        1.00 &         1.00 &          1.00 &           1.00 &          1.00 &      1.00 &       1.00 &        1.00 &         1.00 &          1.00 &         1.00 &       1.00 &        1.00 &         1.00 &          1.00 &           1.00 &          1.00 \\
DES16E2dcg       &       0.71 &        0.01 &         0.50 &          0.33 &           0.56 &          0.42 &      0.95 &       0.98 &        0.94 &         0.95 &          0.93 &         0.95 &       0.88 &        0.86 &         0.75 &          0.69 &           0.96 &          0.83 \\
DES16E2dch       &       1.00 &        1.00 &         1.00 &          1.00 &           1.00 &          1.00 &      1.00 &       1.00 &        1.00 &         1.00 &          1.00 &         1.00 &       1.00 &        1.00 &         1.00 &          1.00 &           1.00 &          1.00 \\
DES15C2iuv       &       1.00 &        1.00 &         1.00 &          1.00 &           1.00 &          1.00 &      1.00 &       1.00 &        1.00 &         1.00 &          1.00 &         1.00 &       1.00 &        1.00 &         1.00 &          1.00 &           1.00 &          1.00 \\
DES13X2gnl       &       1.00 &        1.00 &         1.00 &          1.00 &           1.00 &          1.00 &      1.00 &       1.00 &        1.00 &         1.00 &          1.00 &         1.00 &       1.00 &        1.00 &         1.00 &          1.00 &           1.00 &          1.00 \\
DES16E1byy       &       1.00 &        1.00 &         1.00 &          1.00 &           1.00 &          1.00 &      1.00 &       1.00 &        1.00 &         1.00 &          1.00 &         1.00 &       1.00 &        1.00 &         1.00 &          1.00 &           1.00 &          1.00 \\
DES15E2mhj       &       1.00 &        1.00 &         1.00 &          1.00 &           1.00 &          1.00 &      1.00 &       1.00 &        1.00 &         1.00 &          1.00 &         1.00 &       1.00 &        1.00 &         1.00 &          1.00 &           1.00 &          1.00 \\
DES14E2cmo       &       1.00 &        1.00 &         1.00 &          1.00 &           1.00 &          1.00 &      1.00 &       1.00 &        1.00 &         1.00 &          1.00 &         1.00 &       1.00 &        1.00 &         1.00 &          1.00 &           1.00 &          1.00 \\
DES14E2hhu       &       1.00 &        1.00 &         1.00 &          1.00 &           1.00 &          1.00 &      1.00 &       1.00 &        1.00 &         1.00 &          1.00 &         1.00 &       1.00 &        1.00 &         1.00 &          1.00 &           1.00 &          1.00 \\
DES13E1aftw      &       1.00 &        1.00 &         1.00 &          1.00 &           1.00 &          1.00 &      1.00 &       1.00 &        1.00 &         1.00 &          1.00 &         1.00 &       1.00 &        1.00 &         1.00 &          1.00 &           1.00 &          1.00 \\
DES14C2ocp       &       1.00 &        1.00 &         1.00 &          0.98 &           1.00 &          0.99 &      0.99 &       1.00 &        0.99 &         1.00 &          0.99 &         0.99 &       1.00 &        0.99 &         0.83 &          0.99 &           0.95 &          0.95 \\
DES13X2jdk       &       1.00 &        0.97 &         0.99 &          0.99 &           0.99 &          0.99 &      0.99 &       0.97 &        0.99 &         0.98 &          0.96 &         0.98 &       0.98 &        0.98 &         0.96 &          0.95 &           0.98 &          0.97 \\
DES14E2clm       &       1.00 &        1.00 &         1.00 &          1.00 &           1.00 &          1.00 &      1.00 &       1.00 &        1.00 &         1.00 &          1.00 &         1.00 &       1.00 &        1.00 &         1.00 &          1.00 &           1.00 &          1.00 \\
DES17C3blq       &       0.01 &        1.00 &         0.96 &          0.32 &           0.99 &          0.66 &      0.94 &       1.00 &        0.79 &         0.35 &          0.29 &         0.67 &       0.93 &        0.75 &         0.96 &          0.96 &           0.18 &          0.76 \\
DES16E2blm       &       1.00 &        1.00 &         1.00 &          1.00 &           1.00 &          1.00 &      1.00 &       1.00 &        1.00 &         1.00 &          1.00 &         1.00 &       1.00 &        1.00 &         1.00 &          1.00 &           1.00 &          1.00 \\
DES15X3auw       &       1.00 &        1.00 &         1.00 &          1.00 &           1.00 &          1.00 &      1.00 &       1.00 &        1.00 &         1.00 &          1.00 &         1.00 &       1.00 &        1.00 &         1.00 &          1.00 &           1.00 &          1.00 \\
DES15X2mey       &       1.00 &        1.00 &         1.00 &          1.00 &           1.00 &          1.00 &      1.00 &       1.00 &        1.00 &         1.00 &          1.00 &         1.00 &       1.00 &        1.00 &         1.00 &          1.00 &           1.00 &          1.00 \\
DES15C3odz       &       1.00 &        1.00 &         1.00 &          1.00 &           1.00 &          1.00 &      1.00 &       1.00 &        1.00 &         1.00 &          1.00 &         1.00 &       1.00 &        1.00 &         1.00 &          1.00 &           1.00 &          1.00 \\
DES14C3oce       &       1.00 &        1.00 &         1.00 &          1.00 &           1.00 &          1.00 &      1.00 &       1.00 &        1.00 &         1.00 &          1.00 &         1.00 &       0.73 &        0.99 &         0.89 &          0.99 &           0.98 &          0.92 \\
DES15X2mfa       &       0.99 &        0.99 &         0.99 &          0.89 &           0.99 &          0.97 &      0.99 &       0.97 &        0.97 &         0.94 &          0.95 &         0.96 &       0.93 &        0.98 &         0.94 &          0.93 &           0.96 &          0.95 \\
DES16S2buz       &       1.00 &        1.00 &         1.00 &          1.00 &           1.00 &          1.00 &      1.00 &       1.00 &        1.00 &         1.00 &          1.00 &         1.00 &       1.00 &        1.00 &         1.00 &          1.00 &           1.00 &          1.00 \\
DES16C3fhz       &       1.00 &        1.00 &         1.00 &          1.00 &           1.00 &          1.00 &      1.00 &       1.00 &        1.00 &         1.00 &          1.00 &         1.00 &       1.00 &        1.00 &         1.00 &          1.00 &           1.00 &          1.00 \\
DES17C2emh       &       1.00 &        1.00 &         1.00 &          1.00 &           1.00 &          1.00 &      1.00 &       1.00 &        1.00 &         1.00 &          1.00 &         1.00 &       1.00 &        1.00 &         1.00 &          1.00 &           1.00 &          1.00 \\
DES17E1bmf       &       1.00 &        1.00 &         1.00 &          1.00 &           1.00 &          1.00 &      1.00 &       1.00 &        1.00 &         1.00 &          1.00 &         1.00 &       1.00 &        1.00 &         1.00 &          1.00 &           1.00 &          1.00 \\
DES17C3ivv       &       0.87 &        1.00 &         0.63 &          0.96 &           1.00 &          0.89 &      0.81 &       0.99 &        0.98 &         0.92 &          0.90 &         0.92 &       0.99 &        0.99 &         0.99 &          0.98 &           1.00 &          0.99 \\
DES17E1blu       &       1.00 &        1.00 &         1.00 &          1.00 &           1.00 &          1.00 &      1.00 &       1.00 &        1.00 &         1.00 &          1.00 &         1.00 &       0.99 &        0.97 &         1.00 &          0.98 &           0.98 &          0.98 \\
DES14S2frj       &       1.00 &        1.00 &         1.00 &          1.00 &           1.00 &          1.00 &      1.00 &       0.99 &        1.00 &         1.00 &          1.00 &         1.00 &       1.00 &        1.00 &         1.00 &          1.00 &           1.00 &          1.00 \\
DES17E2bmb       &       1.00 &        1.00 &         1.00 &          1.00 &           1.00 &          1.00 &      1.00 &       1.00 &        1.00 &         1.00 &          1.00 &         1.00 &       1.00 &        1.00 &         1.00 &          1.00 &           1.00 &          1.00 \\
DES17C1ify       &       1.00 &        1.00 &         1.00 &          1.00 &           1.00 &          1.00 &      1.00 &       1.00 &        1.00 &         1.00 &          1.00 &         1.00 &       1.00 &        1.00 &         1.00 &          1.00 &           1.00 &          1.00 \\
DES16X1drk       &       1.00 &        1.00 &         1.00 &          1.00 &           1.00 &          1.00 &      1.00 &       1.00 &        1.00 &         1.00 &          1.00 &         1.00 &       1.00 &        1.00 &         1.00 &          1.00 &           1.00 &          1.00 \\
\end{tabular}
\end{table*}

\section{Author Affiliations}\label{appendix:affiliations}
$^{1}$ Centre for Astrophysics \& Supercomputing, Swinburne University of Technology, Victoria 3122, Australia \\ 
$^{2}$ LPC, Université Clermont Auvergne, CNRS/IN2P3, F-63000 Clermont-Ferrand, France. \\ 
$^{3}$ Université de Lyon, Université Claude Bernard Lyon 1, CNRS/IN2P3, IP2I Lyon, F-69622, Villeurbanne, France. \\ 
$^{4}$ Department of Physics and Astronomy, University of Pennsylvania, Philadelphia, PA 19104, USA \\ 
$^{5}$ School of Physics and Astronomy, University of Southampton,  Southampton, SO17 1BJ, UK \\ 
$^{6}$ Department of Physics, Duke University Durham, NC 27708, USA \\ 
$^{7}$ The Research School of Astronomy and Astrophysics, Australian National University, ACT 2601, Australia \\ 
$^{8}$ Centro de Investigaciones Energ\'eticas, Medioambientales y Tecnol\'ogicas (CIEMAT), Madrid, Spain \\ 
$^{9}$ Center for Astrophysics $\vert$ Harvard \& Smithsonian, 60 Garden Street, Cambridge, MA 02138, USA \\ 
$^{10}$ INAF-Osservatorio Astronomico di Trieste via Tiepolo 11, I-34143 Trieste, Italy. \\ 
$^{11}$ School of Mathematics and Physics, University of Queensland,  Brisbane, QLD 4072, Australia \\ 
$^{12}$ Institute of Cosmology and Gravitation, University of Portsmouth, Portsmouth, PO1 3FX, UK \\ 
$^{13}$ Institute of Space Sciences (ICE, CSIC), Campus UAB, Carrer de Can Magrans, s/n, E-08193 Barcelona, Spain. \\ 
$^{14}$ Institut d'Estudis Espacials de Catalunya (IEEC), E-08034 Barcelona, Spain. \\ 
$^{15}$ Department of Astronomy and Astrophysics, University of Chicago, Chicago, IL 60637, USA \\ 
$^{16}$ Kavli Institute for Cosmological Physics, University of Chicago, Chicago, IL 60637, USA \\ 
$^{17}$ Sydney Institute for Astronomy, School of Physics, A28, The University of Sydney, NSW 2006, Australia \\ 
$^{18}$ Centre for Gravitational Astrophysics, College of Science, The Australian National University, ACT 2601, Australia \\ 
$^{19}$ Department of Physics, University of Surrey, Guildford, Surrey, UK, GU2 7XH \\ 
$^{20}$ Cerro Tololo Inter-American Observatory, NSF's National Optical-Infrared Astronomy Research Laboratory, Casilla 603, La Serena, Chile \\ 
$^{21}$ Laborat\'orio Interinstitucional de e-Astronomia - LIneA, Rua Gal. Jos\'e Cristino 77, Rio de Janeiro, RJ - 20921-400, Brazil \\ 
$^{22}$ Fermi National Accelerator Laboratory, P. O. Box 500, Batavia, IL 60510, USA \\ 
$^{23}$ CNRS, UMR 7095, Institut d'Astrophysique de Paris, F-75014, Paris, France \\ 
$^{24}$ Sorbonne Universit\'es, UPMC Univ Paris 06, UMR 7095, Institut d'Astrophysique de Paris, F-75014, Paris, France \\ 
$^{25}$ Faculty of Physics, Ludwig-Maximilians-Universit\"at, Scheinerstr. 1, 81679 Munich, Germany \\ 
$^{26}$ Department of Physics \& Astronomy, University College London, Gower Street, London, WC1E 6BT, UK \\ 
$^{27}$ Kavli Institute for Particle Astrophysics \& Cosmology, P. O. Box 2450, Stanford University, Stanford, CA 94305, USA \\ 
$^{28}$ SLAC National Accelerator Laboratory, Menlo Park, CA 94025, USA \\ 
$^{29}$ Instituto de Astrofisica de Canarias, E-38205 La Laguna, Tenerife, Spain \\ 
$^{30}$ Universidad de La Laguna, Dpto. Astrofísica, E-38206 La Laguna, Tenerife, Spain \\ 
$^{31}$ Center for Astrophysical Surveys, National Center for Supercomputing Applications, 1205 West Clark St., Urbana, IL 61801, USA \\ 
$^{32}$ Department of Astronomy, University of Illinois at Urbana-Champaign, 1002 W. Green Street, Urbana, IL 61801, USA \\ 
$^{33}$ Institut de F\'{\i}sica d'Altes Energies (IFAE), The Barcelona Institute of Science and Technology, Campus UAB, 08193 Bellaterra (Barcelona) Spain \\ 
$^{34}$ Institut d'Estudis Espacials de Catalunya (IEEC), 08034 Barcelona, Spain \\ 
$^{35}$ Institute of Space Sciences (ICE, CSIC),  Campus UAB, Carrer de Can Magrans, s/n,  08193 Barcelona, Spain \\ 
$^{36}$ Jodrell Bank Center for Astrophysics, School of Physics and Astronomy, University of Manchester, Oxford Road, Manchester, M13 9PL, UK \\ 
$^{37}$ University of Nottingham, School of Physics and Astronomy, Nottingham NG7 2RD, UK \\ 
$^{38}$ Astronomy Unit, Department of Physics, University of Trieste, via Tiepolo 11, I-34131 Trieste, Italy \\ 
$^{39}$ INAF-Osservatorio Astronomico di Trieste, via G. B. Tiepolo 11, I-34143 Trieste, Italy \\ 
$^{40}$ Institute for Fundamental Physics of the Universe, Via Beirut 2, 34014 Trieste, Italy \\ 
$^{41}$ Observat\'orio Nacional, Rua Gal. Jos\'e Cristino 77, Rio de Janeiro, RJ - 20921-400, Brazil \\ 
$^{42}$ Department of Physics, IIT Hyderabad, Kandi, Telangana 502285, India \\ 
$^{43}$ Santa Cruz Institute for Particle Physics, Santa Cruz, CA 95064, USA \\ 
$^{44}$ Institute of Theoretical Astrophysics, University of Oslo. P.O. Box 1029 Blindern, NO-0315 Oslo, Norway \\ 
$^{45}$ Instituto de Fisica Teorica UAM/CSIC, Universidad Autonoma de Madrid, 28049 Madrid, Spain \\ 
$^{46}$ Department of Astronomy, University of Michigan, Ann Arbor, MI 48109, USA \\ 
$^{47}$ Department of Physics, University of Michigan, Ann Arbor, MI 48109, USA \\ 
$^{48}$ Excellence Cluster Origins, Boltzmannstr.\ 2, 85748 Garching, Germany \\ 
$^{49}$ Center for Cosmology and Astro-Particle Physics, The Ohio State University, Columbus, OH 43210, USA \\ 
$^{50}$ Department of Physics, The Ohio State University, Columbus, OH 43210, USA \\ 
$^{51}$ Australian Astronomical Optics, Macquarie University, North Ryde, NSW 2113, Australia \\ 
$^{52}$ Lowell Observatory, 1400 Mars Hill Rd, Flagstaff, AZ 86001, USA \\ 
$^{53}$ George P. and Cynthia Woods Mitchell Institute for Fundamental Physics and Astronomy, and Department of Physics and Astronomy, Texas A\&M University, College Station, TX 77843,  USA \\ 
$^{54}$ Instituci\'o Catalana de Recerca i Estudis Avan\c{c}ats, E-08010 Barcelona, Spain \\ 
$^{55}$ Physics Department, 2320 Chamberlin Hall, University of Wisconsin-Madison, 1150 University Avenue Madison, WI  53706-1390 \\ 
$^{56}$ Department of Astronomy, University of California, Berkeley,  501 Campbell Hall, Berkeley, CA 94720, USA \\ 
$^{57}$ Institute of Astronomy, University of Cambridge, Madingley Road, Cambridge CB3 0HA, UK \\ 
$^{58}$ Department of Astrophysical Sciences, Princeton University, Peyton Hall, Princeton, NJ 08544, USA \\ 
$^{59}$ Department of Physics and Astronomy, Pevensey Building, University of Sussex, Brighton, BN1 9QH, UK \\ 
$^{60}$ Computer Science and Mathematics Division, Oak Ridge National Laboratory, Oak Ridge, TN 37831 \\ 
$^{61}$ Max Planck Institute for Extraterrestrial Physics, Giessenbachstrasse, 85748 Garching, Germany \\ 
$^{62}$ Universit\"ats-Sternwarte, Fakult\"at f\"ur Physik, Ludwig-Maximilians Universit\"at M\"unchen, Scheinerstr. 1, 81679 M\"unchen, Germany


\bsp    
\label{lastpage}
\end{document}